\documentclass[useAMS,usenatbib]{mn2e}
\pdfoutput=1
\usepackage{amsfonts,amsmath,amssymb}
\usepackage{txfonts}
\usepackage{graphicx}
\usepackage{color}
\usepackage{flushend}
\usepackage{enumitem}
\usepackage{tabularx}
\usepackage[breakwithin,restart,roman]{parnotes}
\usepackage[colorlinks,linkcolor=blue,citecolor=blue,urlcolor=blue,pdftitle={Mass and galaxy distributions of four massive galaxy clusters from Dark Energy Survey Science Verification data},pdfauthor={Melchior et al.}]{hyperref}
\renewcommand{\parnotefmt}[1]{\footnotesize\raggedright #1}

\DeclareRobustCommand{\object}[1]{%
   #1%
}
\newcommand\rxj{\object{RXC J2248.7-4431}}
\newcommand\bulletcl{\object{1E 0657-56}}
\newcommand\sptw{\object{SCSO J233227-535827}}
\newcommand\abell{\object{Abell 3261}}

\newcommand\sqdegree{deg$^2$}
\newcommand\shape{{\sc im3shape}}
\newcommand\balrog{{\sc Balrog}}
\newcommand\sextractor{{\sc SExtractor}}
\newcommand\psfex{{\sc PSFEx}}
\newcommand\swarp{{\sc SWarp}}
\newcommand\scamp{{\sc scamp}}
\newcommand\bigmacs{{\sc Big~MACS}}
\newcommand\healpix{{\sc HEALPix}}
\newcommand\sersic{S\'{e}rsic}
\newcommand\redmapper{{redMaPPer}}
\newcommand{\zlambda}{z_{\lambda}}
\newcommand{\photoz}{photo-$z$}
\newcommand\conj[1]{#1^*}

\title[Mass and galaxy distributions of four massive galaxy clusters with DES]{Mass and galaxy distributions of four massive galaxy clusters from Dark Energy Survey Science Verification data}
\author[Melchior et al.]{%
\parbox{\textwidth}{\raggedright
\mbox{P. Melchior$^{1,2}$\thanks{E-mail: melchior.12@osu.edu}},
\mbox{E. Suchyta$^2$},
\mbox{E. Huff$^1$},
\mbox{M. Hirsch$^3$},
\mbox{T. Kacprzak$^4$},
\mbox{E. Rykoff$^5$},
\mbox{D. Gruen$^{6,7}$},
\mbox{R. Armstrong$^8$},
\mbox{D. Bacon$^{9}$}, 
\mbox{K. Bechtol$^{10}$}, 
\mbox{G. M. Bernstein$^8$},
\mbox{S. Bridle$^4$},
\mbox{J. Clampitt$^8$},
\mbox{K. Honscheid$^2$}, 
\mbox{B. Jain$^8$},
\mbox{S. Jouvel$^3$},
\mbox{E. Krause$^8$}, 
\mbox{H. Lin$^{11}$}, 
\mbox{N. MacCrann$^4$},
\mbox{K. Patton$^2$},
\mbox{A. Plazas$^{12}$},
\mbox{B. Rowe$^3$},
\mbox{V. Vikram$^8$}, 
\mbox{H. Wilcox$^{9}$},
\mbox{J. Young$^6$},
\mbox{J. Zuntz$^4$}, 
\mbox{T. Abbott$^{13}$},
\mbox{F. B. Abdalla$^3$},
\mbox{S. S. Allam$^{14,11}$},
\mbox{M. Banerji$^3$},
\mbox{J. P. Bernstein$^{15}$},
\mbox{R. A. Bernstein$^{16}$},
\mbox{E. Bertin$^{17}$},
\mbox{E. Buckley-Geer$^{11}$},
\mbox{D. L. Burke$^{5,18}$},
\mbox{F. J. Castander$^{19}$},
\mbox{L. N. da Costa$^{20,21}$},
\mbox{C. E. Cunha$^{18}$},
\mbox{D. L. Depoy$^{22}$},
\mbox{S. Desai$^{23,24}$},
\mbox{H. T. Diehl$^{11}$},
\mbox{P. Doel$^3$},
\mbox{J. Estrada$^{11}$},
\mbox{A. E. Evrard$^{25,26,17}$},
\mbox{A. Fausti Neto$^{21}$},
\mbox{E. Fernandez$^{27}$},
\mbox{D. A. Finley$^{11}$},
\mbox{B. Flaugher$^{11}$},
\mbox{J. A. Frieman$^{11,10}$},
\mbox{E. Gaztanaga$^{19}$},
\mbox{D. Gerdes$^{25}$},
\mbox{R. A. Gruendl$^{28,29}$},
\mbox{G. R. Gutierrez$^{11}$},
\mbox{M. Jarvis$^8$},
\mbox{I. Karliner$^{30}$},
\mbox{S. Kent$^{11}$},
\mbox{K. Kuehn$^{31}$},
\mbox{N. Kuropatkin$^{11}$},
\mbox{O. Lahav$^3$},
\mbox{M. A. G. Maia$^{20,21}$},
\mbox{M. Makler$^{32}$},
\mbox{J. Marriner$^{11}$},
\mbox{J. L. Marshall$^{22}$},
\mbox{K. W. Merritt$^{11}$},
\mbox{C. J. Miller$^{26}$},
\mbox{R. Miquel$^{27,33}$},
\mbox{J. Mohr$^{23}$},
\mbox{E. Neilsen$^{11}$},
\mbox{R. C. Nichol$^9$},
\mbox{B. D. Nord$^{11}$},
\mbox{K. Reil$^{5,18}$},
\mbox{N. A. Roe$^{34}$},
\mbox{A. Roodman$^{5,18}$},
\mbox{M. Sako$^8$},
\mbox{E. Sanchez$^{35}$},
\mbox{B. X. Santiago$^{36,21}$},
\mbox{R. Schindler$^{5,18}$},
\mbox{M. Schubnell$^{25}$},
\mbox{I. Sevilla-Noarbe$^{35}$},
\mbox{E. Sheldon$^{12}$},
\mbox{C. Smith$^{13}$},
\mbox{M. Soares-Santos$^{11}$},
\mbox{M. E. C. Swanson$^{29}$},
\mbox{A. J. Sypniewski$^{25}$},
\mbox{G. Tarle$^{25}$},
\mbox{J. Thaler$^{30}$},
\mbox{D. Thomas$^{9,37}$},
\mbox{D. L. Tucker$^{11}$},
\mbox{A. Walker$^{13}$},
\mbox{R. Wechsler$^{5,18}$},
\mbox{J. Weller$^{6,7,24}$},
\mbox{W. Wester$^{11}$}
\vspace{0.3cm}\\
\parbox{\textwidth}{\small $^\star$ Direct your inquiries to \href{mailto:melchior.12@osu.edu}{melchior.12@osu.edu}. Author affiliations are listed at the end of this paper.
}}}

\begin{document}
\date{}
\pagerange{\pageref{firstpage}--\pageref{lastpage}} \pubyear{2015}
\maketitle
\label{firstpage}

\begin{abstract}
We measure the weak-lensing masses and galaxy distributions of four massive galaxy clusters observed during the Science Verification phase of the Dark Energy Survey. This pathfinder study is meant to 1) validate the DECam imager for the task of measuring weak-lensing shapes, and 2) utilize DECam's large field of view to map out the clusters and their environments over 90 arcmin. We conduct a series of rigorous tests on astrometry, photometry, image quality, PSF modeling, and shear measurement accuracy to single out flaws in the data and also to identify the optimal data processing steps and parameters. 
We find Science Verification data from DECam to be suitable for the lensing analysis described in this paper. The PSF is generally well-behaved, but the modeling is rendered difficult by a flux-dependent PSF width and ellipticity. We employ photometric redshifts to distinguish between foreground and background galaxies, and a red-sequence cluster finder to provide cluster richness estimates and cluster-galaxy distributions. By fitting NFW profiles to the clusters in this study, we determine weak-lensing masses that are in agreement with previous work. 
For \abell, we provide the first estimates of redshift, weak-lensing mass, and richness. In addition, the cluster-galaxy distributions indicate the presence of filamentary structures attached to \bulletcl{} and \rxj, stretching out as far as 1 degree (approximately 20 Mpc), showcasing the potential of DECam and DES for detailed studies of degree-scale features on the sky.
\end{abstract}


\section{Introduction}

The Dark Energy Survey (DES) comprises an optical to near-infrared survey over 5,000 \sqdegree{} of the South Galactic Cap to $\sim$24th magnitude in the SDSS \emph{grizY} bands and a time-domain \emph{griz} survey over 30 \sqdegree{} with a cadence of approximately six days. These interleaved surveys are being carried out over 525 nights in the course of five years using the 570-megapixel imager DECam \citep{Flaugher12, Diehl12.1} mounted at the prime focus of the Blanco 4m telescope at NOAO's Cerro Tololo Inter-American Observatory. DECam was commissioned in September and October of 2012, followed by an extended testing and survey commissioning period known as DES Science Verification (SV, November 2012 -- February 2013). With this new instrument, DES will go beyond the reach of SDSS by virtue of telescope aperture, median seeing, and CCD sensitivity, particularly towards the infrared part of the spectrum. Consequently, the galaxy redshift distribution is expected to have a median $z\approx 0.7$ and a significant tail beyond $z=1$, which enables DES to detect clusters at high redshift ($z\approx 1$) and to use source galaxies for a rigorous lensing analysis of clusters beyond $z\approx 0.5$.  DES will also exceed deep and medium-deep weak lensing surveys (CFHTLS, RCS2, DLS) by up to an order of magnitude in area. More details about the survey can be found in \citet{DES05.1} and at \url{http://www.darkenergysurvey.org}.

The very wide field of view of DECam of slightly more than 3 \sqdegree{} allows us to capture the environment of even the most massive galaxy clusters with a single pointing. In this paper we will show results on four fields containing clusters with masses of approximately $1-2\cdot 10^{15}\ \mathrm{M}_\odot$ and redshifts from $z=0.22$ to $z=0.40$. They have been chosen to demonstrate the capabilities of DECam imaging for cluster and lensing analyses and provide an outlook of the utility of the entire DES to map out a good fraction of the sky to redshifts of about 1.
 
The goals of this pathfinder study are twofold. First, to validate the data quality delivered by DECam for the purpose of galaxy cluster and lensing studies. We focus our attention on four fields imaged during the SV period in $grizY$ filters with integration times characteristic of the DES to study the relevant elements of photometry and image quality. We inspect our ability to model the point spread function (PSF) and to account for possible systematic contaminants of photometry and lensing analyses. We emphasize that this is not a comprehensive study of the DES pipelines for photometry or lensing; those studies will be presented elsewhere.

Second, we want to utilize the large field of view of DECam to map the environments of these clusters over 90 arcmin, probing cluster-centric distances between 10 and 15 Mpc at the respective cluster redshifts. We select background galaxies according to their \photoz{} estimates, use the red-sequence cluster-finder \redmapper{} to identify cluster galaxies, and employ \shape{} for weak-lensing shape measurements, from which we obtain mass estimates and two-dimensional mass maps.

\autoref{sec:data} contains the details of the observations and data reduction pipeline used in this study. \autoref{sec:photo} describes the photometric calibration (\autoref{sec:photo_calib}), the \redmapper{} technique for identifying red-sequence galaxies (\autoref{sec:cl_sel}), the photometric redshift methodology (\autoref{sec:photo-z}), and the background galaxy selection procedure (\autoref{sec:bg_sel}). \autoref{sec:shape} describes the lensing analysis, detailing PSF estimation (\autoref{sec:psfmodels}), shape measurements with \shape{} (\autoref{sec:shapes}), and the combination of measurements in three bands into a single shape catalog (\autoref{sec:shape_comb}). We perform additional tests of the recovered cluster shears in \autoref{sec:shear_sys} and present the NFW-profile fits and lensing mass estimates in \autoref{sec:masses} and the mass-richness relation in \autoref{sec:mass-richness}. We show mass and  cluster galaxy maps in \autoref{sec:maps} and indications for the presence of filamentary structures in two of the investigated fields (\autoref{sec:filaments}). We summarize our findings in \autoref{sec:discussion}.

For the entire paper we adopt a flat $\Lambda$CDM cosmological model with $\Omega_m$ = 0.3 and $H_0 = 100\ h$ km/s/Mpc, where $h = 0.7$.

\section{Observations and data processing}
\label{sec:data}

\begin{table}
\caption{The cluster sample.  Coordinates correspond to the centroids of the Brightest Cluster Galaxies (BCGs), redshifts are spectroscopically confirmed. The labels are used as abbreviations throughout this work.}
\setlength{\tabcolsep}{.5em}
\label{tab:clusters}
\begin{tabularx}{\linewidth}{lllll}
Cluster name & Label & RA [deg] & Dec [deg] & $z$\\
\hline
\rxj & RXJ & $342.18319$ & $-44.53091$ & $0.348$\\
\bulletcl & Bullet & $104.64708$ & $-55.94897$ & $0.296$\\
\sptw & SPTW1 & $353.11446$ & $-53.97441$ & $0.402$\\
\abell & Abell3261 & $67.31375$ & $-60.32555$ & ---\\
\hline
\end{tabularx}
\end{table}

The clusters targeted for this study are both massive and at intermediate redshift so as to show up prominently in our weak-lensing measurements. In detail, we targeted \bulletcl{} \citep[known as the Bullet cluster]{Tucker98.1}, \rxj{} \citep{Boehringer04.1}, and \sptw{} \citep{Menanteau10.1}.\footnote{In the course of this program, we also observed ACT-CL J0102-4915 \citep[dubbed El Gordo,][]{Menanteau10.2}, but the images are rather shallow for a dedicated weak-lensing analysis of this cluster at $z=0.87$. We therefore omit the cluster in this work.} All of these systems are well studied, providing us with important information such as spectroscopic redshifts and mass estimates from lensing or baryonic tracers. General properties of the clusters are listed in \autoref{tab:clusters}.

The exposures for these clusters were taken over the course of several nights (November 16 -- 24 and December 7, 2012). We adhered to the nominal DES exposure times: 90 seconds in $g,r,i,z$ and 50 seconds in $Y$, and used a 10-exposure dither pattern centered on the cluster with offsets of around 0.1 deg. Hence, the total depth of these observations is characteristic of the DES main survey, but differs in the dither pattern. 

We extended the data set in two ways. First, we re-observed one
cluster,  \rxj, later in the season (on August 15, 2013) to benefit
from improvements to telescope performance and
general image quality. Second, in order to compare our findings from
the targeted cluster fields to typical DES survey performance, we
added another cluster, \abell{} \citep{Abell89.1},  to our
investigation. It was observed during tests of survey operations, when DES observations are coordinated by an automated observer tactician program, {\sc ObsTac} \citep{Neilsen13.1}, which selects upcoming exposures based upon survey history and current conditions\footnote{One key decision made by {\sc ObsTac} is to observe in the $r,i,z$ bands only if the seeing in the proposed band is better than $1\farcs1$, taking into account the chromatic and airmass dependence.}. Thus, for this cluster the imaging data experience seeing conditions, sky brightness levels, and the dither pattern typical of the main survey.

All raw data presented here, including the calibration images, are public and can be obtained from the \href{http://portal-nvo.noao.edu/}{NOAO archive}.

\subsection{Observational conditions}
\label{sec:observations}
The conditions during the observations were generally stable and mostly photometric. 
During the November runs, the moon illumination was bright enough to significantly reduce the depth in $g$ and $r$, triggering re-observations during dark conditions on December 7.
The seeing in these nights varied between $0\farcs8$ and $1\farcs0$ in $i$, with larger seeing values in $r$ and particularly $g$, in agreement with atmospheric turbulence models. The stellar ellipticity was typically smooth across the field of view\footnote{to the degree we could determine from our observations, which comprise the Bullet cluster field that exhibits a strongly enhanced stellar density due to its low galactic latitude. However, dedicated studies on star clusters were not performed.}, and varied only slowly with time during the same night, indicating that the PSF is dominated by the optics rather than the atmosphere. If observations spanned multiple nights, non-trivial variations occurred, which render the PSF modeling more challenging.
The observations from November exhibit a predominant elongation in the right ascension direction due to telescope tracking oscillations. Because of mechanical improvements of the telescope between these first sets of exposures and the re-observations on December 7 and later during the SV period, the latter ones show smaller overall ellipticity and can be well described by the typical optical aberrations of a wide-field imager. An overview of average seeing conditions for all data in question is given in \autoref{tab:seeing}.

\begin{table}
\caption{Average PSF width (seeing) and ellipticity for each filter in
  each of the four fields, with re-observations indicated by a running number. Observations
  marked with $\dagger$ were completed in the following night; those
  marked with $\ddagger$ were taken under \textsc{ObsTac} control
  during survey-mode operations between December 2012 and January
  2013. Seeing is given in terms of the FWHM in arcsec, the ellipticity in terms of shears, not polarizations. The last column 'Shapes' denotes whether the coadd image was suitable for a weak-lensing shape analysis (see \autoref{sec:psftests} for details on the selection).}
\label{tab:seeing}
\begin{tabular}{lllllll}
Field & Band & Date  & Seeing & Ellipticity & Shapes\\
\hline
Bullet & $g$ & 2012-12-07 & 1.06 & 0.038 & \\
Bullet & $r.1$ & 2012-11-23$^\dagger$ & 1.04 & 0.056 & $\checkmark$\\
Bullet & $r.2$ & 2012-12-07 & 0.93 & 0.027 & $\checkmark$\\
Bullet & $i$ & 2012-11-23$^\dagger$ & 1.00 & 0.032 & $\checkmark$\\
Bullet & $z$ & 2012-11-23$^\dagger$ & 0.97 & 0.039 & $\checkmark$\\
Bullet & $Y$ & 2012-11-23$^\dagger$ & 1.00 & 0.047 & \\
\hline
RXJ & $g$ & 2012-12-07 & 1.18 & 0.031 & \\
RXJ & $r.1$ & 2012-11-24 & 0.92 & 0.041 & $\checkmark$\\
RXJ & $r.2$ & 2012-12-07 & 1.08 & 0.019 & $\checkmark$\\
RXJ & $i.1$ & 2012-11-24 & 0.86 & 0.029 & \\
RXJ & $i.2$ & 2013-08-15 & 0.79 & 0.023 & $\checkmark$\\
RXJ & $z.1$ & 2012-11-24 & 0.90 & 0.045 & \\
RXJ & $z.2$ & 2013-08-15 & 0.76 & 0.027 & $\checkmark$\\
RXJ & $Y$ & 2012-11-24 & 0.85 & 0.042 & \\
\hline
SPTW1 & $g.1$ & 2012-11-16$^\dagger$ & 1.4 & 0.025 & \\
SPTW1 & $g.2$ & 2012-12-07 & 1.13 & 0.037 & \\
SPTW1 & $r$ & 2012-11-17 & 0.97 & 0.027 & $\checkmark$\\
SPTW1 & $i$ & 2012-11-18 & 0.99 & 0.036 & $\checkmark$\\
SPTW1 & $z$ & 2012-11-17 & 0.90 & 0.029 & $\checkmark$\\
SPTW1 & $Y$ & 2012-11-16 & 1.15 & 0.029 & \\
\hline
Abell3261 & $g$ & $\ddagger$ & 1.10 & 0.024 & \\
Abell3261 & $r$ & $\ddagger$ & 0.95 & 0.026 & $\checkmark$\\
Abell3261 & $i$ & $\ddagger$ & 0.87 & 0.016 & $\checkmark$\\
Abell3261 & $z$ & $\ddagger$ & 1.03 & 0.024 & $\checkmark$\\
Abell3261 & $Y$ & $\ddagger$ & 0.87 & 0.028 & \\
\hline
\end{tabular}
\end{table}

\subsection{Coadd images and catalogs}
\label{sec:coadds}

Our treatment of the data begins with single-epoch images
that have been processed through the DES Data Management (DESDM)
pipeline \citep{desdm, Desai12.1}.
Because a significant portion of our observations was taken early in the SV program,
we first visually inspected our images for adequate data quality and rejected problematic
exposures, e.g. those affected by occasional guiding failures. We excluded symptomatic frames, characterized by elongated, distorted stellar profiles. In a few instances we also rejected exposures showing significant levels of scattered light across the focal plane.

The DESDM pipeline already implements several standard detrending corrections for the raw data.
The overscan is subtracted and a crosstalk matrix removes
effects that bright stars induce across the detector. Bias frames and dome flats are
averaged over several nights and then applied to the data. 
Included in the DESDM reductions are fringe and pupil-ghost corrections as well as the photometric calibration, which we detail in \autoref{sec:photo_calib}.

Astrometric solutions are computed by DESDM using \scamp{} \citep{scamp}, matching absolute stellar positions in each exposure separately to the
UCAC4 reference catalog \citep{ucac4}. While we do not require precise absolute astrometry,
the relative astrometry between individual exposures is critical for accurate shape measurements.
We therefore re-run \scamp, simultaneously matching all filters and exposures to each other as well as to the much sparser reference catalog.
Typical errors after this step are 5 mas shifts between bands and
20 mas scatter within a band, improving the single-epoch solutions by factors of $\sim3$.

\begin{figure}
\includegraphics[width=\linewidth]{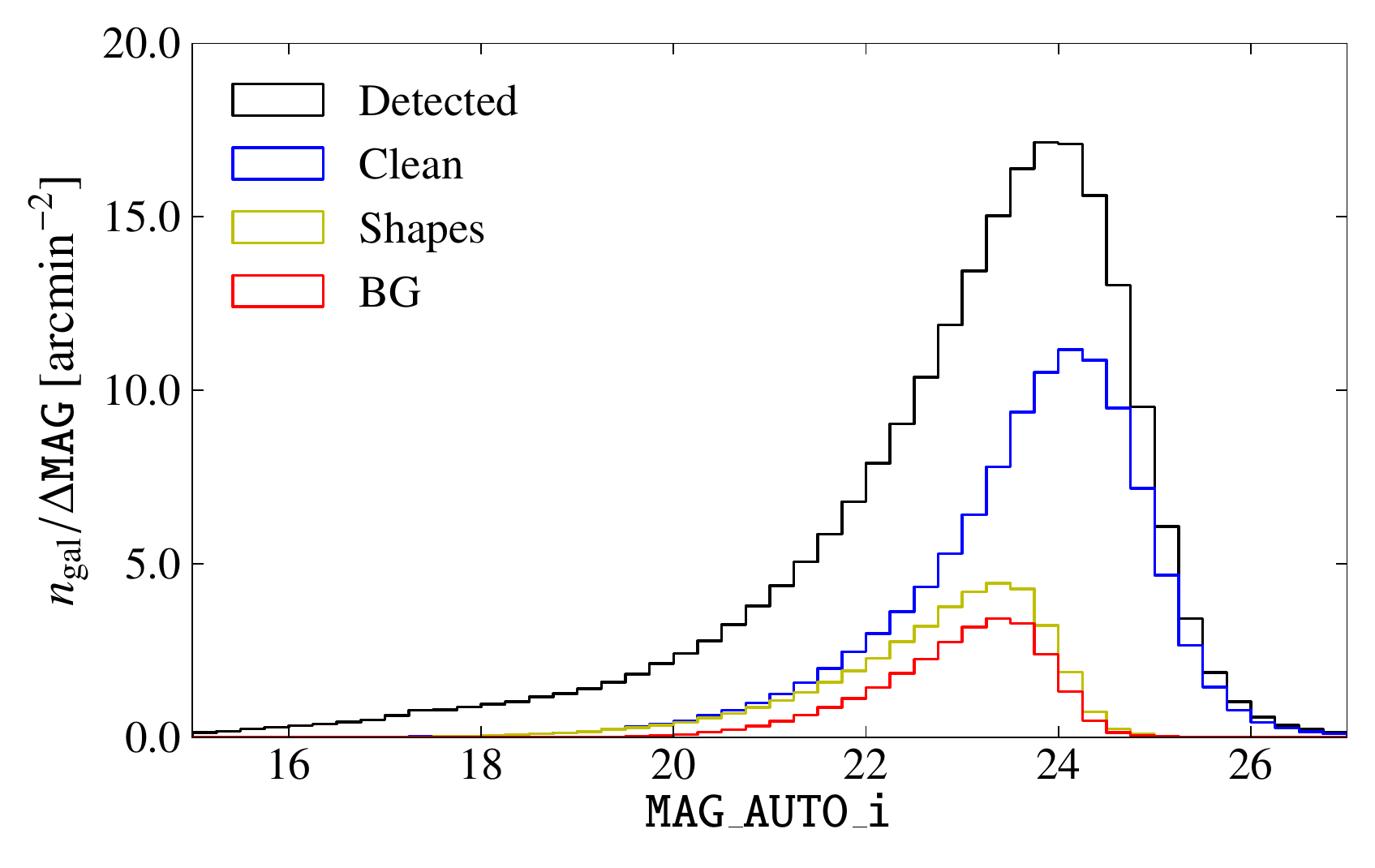}
\caption{Galaxy number density $n_\textrm{gal}$ in bins of \sextractor's magnitude \texttt{MAG\_AUTO} in the $i$-band. From top to bottom: detected sources (\emph{black}); after cleaning the catalog (\emph{blue}); galaxies with successful shape measurements (\emph{yellow}); galaxies considered in the lensed background sample (\emph{red}). See \autoref{sec:shape_comb} for details on the selections. The numbers are unweighted, averaged over all four fields.}
\label{fig:mag_distribution}
\end{figure}

We then coadd the single-epoch images of each $g,r,i,z,Y$ filter band separately.
Additionally, we combine the bands considered for shape measurement into a multi-band $riz$ coadd for use as a source detection image of increased depth and redshift coverage. 
Coaddition is implemented with \swarp{} \citep{swarp}, using a clipped mean 
algorithm we have added to the software \citep{Gruen13.3}, which rejects pixels that are outliers at the $4\sigma$ level.\footnote{In contrast to median coadds, this approach successfully rejects cosmic-ray hits without sacrificing image fidelity and statistical optimality or altering the PSF shape, critical for weak lensing studies.} During coaddition, appropriate flux scalings are applied to each image using the zeropoints calculated by the DESDM photometric calibrations.

We run \sextractor{} \citep{sextractor} in
\emph{dual-image mode}: we provide the $riz$ coadd for detection of sources and assignment of pixels to detections, afterwards the photometric measurements are made from the single-band values in these pixels.
The detection threshold is set to {$1.5 \sigma$} above noise in at least six contiguous pixels. The magnitude distribution of detected sources can be seen in \autoref{fig:mag_distribution}.

Finally, we define a field mask to restrict the analysis to full-depth areas, excluding the edges of each pointed field (RXJ, Bullet, and SPTW1). We decided on a square shape with a side-length of 1.5 deg with trimmed corners (shown in the left two panels of \autoref{fig:psfmodel}), the center of which was adjusted so as to maximize the number of total detections. 

\section{Photometry and galaxy sample selection}
\label{sec:photo}

In this work, the major role of photometry is to separate galaxies by color and to assign photometric redshifts to them for proper calculation of cosmological distances and lensing weight factors (\autoref{sec:shape_comb}). We will therefore only introduce the general photometric approach, focusing on aspects most relevant for this study and discussing the accuracy with which we can perform these photometric tasks, and refer the reader to forthcoming publications for details.

\subsection{Photometric calibration}
\label{sec:photo_calib}

Each individual exposure is first processed to bring all pixels of the array onto a common photometric scale. Details are given in \citet{Tucker07.1}, so we only summarize the main aspects here. After bias subtraction, the images are divided by a dome flat. The dome flat is known, however, to have variations due to stray light and to changes in the effective pixel area.  To reduce the impact of these variations on the object photometry we divide the images by a \emph{star flat} \citep[e.g.][]{Manfroid95.1}, which provides a correction for $512\times512$ pixel regions on each CCD (32 regions per CCD) that minimizes the dependence of the stellar photometry on the DECam focal plane position.

Exposures taken on clear nights are brought to a common absolute
magnitude scale by using a zeropoint that is a linear function of
airmass, with slope and offset fit to observations of SDSS photometric
standard fields taken at the beginning and end of the night.  At
least one photometric exposure is required for each filter in each
field.  We use these observations to create a local set of standard
stars, for each field, by averaging over all overlapping objects.  We
then refit all CCD zeropoints from both photometric and
non-photometric conditions by allowing each DECam CCD on each exposure
to have its own zeropoint.  Zeropoints are adjusted to minimize the
difference in magnitude between observations of common objects in
different exposures and observations of the aforementioned local
standards.  This method typically gives an RMS accuracy of $\sim1\%$
and has been validated by comparing measurements of the stellar locus
in color-color space, as described below.

Stellar locus regression (SLR) has proven to be a useful complementary photometric
calibration method to standard star observations, and relies upon the approximate
universality of the intrinsic colors of Milky Way halo stars as
a population \citep[e.g.,][]{High09.1}. In the SLR approach,
zeropoints for each filter are adjusted until the foreground stars lie along
the expected color-color locus.
Since the great majority of stars detected in our DECam images are located
beyond the Galactic dust sheet,
the SLR is sensitive to the combined effects of atmospheric and
interstellar extinction.

Our SLR implementation employs the publically available
\bigmacs\footnote{\url{http://code.google.com/p/big-macs-calibrate/}}
code developed by \cite{Kelly12.1}. The reference stellar locus is
synthetically generated using the \cite{Pickles98.1} stellar
spectroscopic library spliced with SDSS spectra \citep[as described in][]{Kelly12.1}
and convolved with the DECam total system transmission functions. Stars
in the cluster fields are selected by requiring \sextractor's stellarity parameter \verb|CLASS_STAR| $>0.95$ in
both the $r$ and $i$ band, which provides adequate star-galaxy
separation for high signal-to-noise objects. Accordingly, we also require magnitude uncertainties of
$<0.05$ ($<0.1$) in the $ri$ ($gz$) bands. In order to evaluate the photometric
calibration across each cluster field, stars are binned into 
\healpix \footnote{\url{http://healpix.sourceforge.net/}} \citep{Gorski05.1}
pixels with a resolution of $\sim$14 arcmin (\verb|NSIDE=256|).\footnote{The choice of 14 arcmin constitutes a compromise between increasing angular resolution and maintaining a sufficient number of stars in each pixel such that statistical uncertainties are $\sim0.02$ magnitudes.}
We then allow the zeropoints of the DECam
$griz$ filters to float independently in each spatial pixel during the
fits and use the $J$-band magnitudes of matched 2MASS stars for
absolute calibration. The zeropoint shifts fitted via SLR are
typically $\lesssim$0.05 mag, with an associated statistical precision of
$\sim$0.02 mag estimated via bootstrapping of the stellar sample. The exception
is the Bullet cluster field, where the median $g$-band zeropoint shift is
$\sim$0.2 mag, consistent with interstellar extinction expected in
this low-Galactic-latitude field from the dust maps produced by \cite{Schlegel98.1}.

\subsection{Cluster member selection}
\label{sec:cl_sel}

Our study uses the methodology from the red-sequence matched filter
probabilistic percolation (``\redmapper'') algorithm~\citep{rykoff+13}, based
on the optimized richness estimator $\lambda$~\citep{rykoff+12}.  \redmapper{}
is a photometric cluster finder that identifies galaxy clusters as
overdensities of red-sequence galaxies, and has been shown to have excellent
performance in \photoz{} determination, purity, and completeness for wide-field
photometric surveys~\citep{rozorykoff13}.  The algorithm is divided into two
stages: the first is a calibration stage where the red-sequence model is
derived directly from the data, and the second is the cluster-finding stage.
These two stages are iterated several times before a final cluster-finding run
is performed.

In the calibration phase, \redmapper{} empirically calibrates the color
distribution (mean and scatter) of red-sequence galaxies as a function of
redshift and magnitude.  For the red-sequence calibration, 
356 spectroscopically confirmed BCGs in the 241-deg$^2$ $griz$ DES SV-A1 galaxy catalog (Rykoff et~al., in prep.) were used. The spectroscopic redshifts were taken from SDSS DR10 \citep{Ahn14.1}, SPT clusters \citep{High10.1}, and as part of the OzDES program (Lidman et al., in prep). These galaxies are used as ``seeds'' to look for
significant overdensities of nearby galaxies with similar color as the seed galaxy ($g-r$, $r-i$,
or $i-z$ depending on the redshift, as determined from \verb|MAG_DETMODEL|\footnote{In \sextractor's dual-image mode, \texttt{MAG\_DETMODEL} measures the flux in each filter by adopting a model fit to the object in the detection image, in our case the $riz$ coadd.} magnitudes).  The resulting set of cluster galaxies is used
to fit a full red-sequence model including zero point, tilt, and scatter.
This scatter is characterized by a full covariance matrix among all colors
(see \citet{rykoff+13} for details). The red-sequence model is calibrated down to a
luminosity threshold of $0.2\,L_*$ at the cluster redshift, which was
determined to be the optimal depth for cluster richness
estimation \citep{rykoff+12}.  In this way we leverage the bright spectroscopic
sample to obtain a model of the red sequence that extends to faint
magnitudes.

Given the red-sequence model and the corrected magnitudes, the cluster-finding
proceeds as follows.  First, we consider all photometric galaxies as candidate
cluster centers.  The red-sequence model is used to calculate a photometric
redshift for each galaxy, and evaluate the goodness of fit of our red-sequence
template.  Galaxies that are not a reasonable fit to the model at any redshift
are immediately discarded.  For the remaining galaxies, we use this initial
redshift guess to evaluate the richness $\lambda$ and the total cluster likelihood.
When at least 3 red sequence galaxies (brighter than $0.2\,L_*$ within a scale
radius $r_\lambda$) are detected, we re-estimate the cluster
redshift by performing a simultaneous fit of all the high-probability cluster
members to the red-sequence model.  This procedure is iterated until
convergence is achieved between member selection and cluster photometric
redshift, denoted $\zlambda$.  The resulting list of candidate cluster centers
is then rank-ordered according to likelihood, and membership probabilities are
used to mask out member galaxies in the final percolation step.  
All richness values are corrected for variations in the local depth of the DES imaging
(Rykoff et al., in prep.), however, the DES imaging is deep enough that this has a 
negligible effect at the redshifts of the clusters considered in this paper. 
We have shown
that for $\lambda>20$ the cluster purity and completeness are $>95\%$
\citep{rykoff+13,rozorykoff13}.  In addition, even for $\lambda>5$ the
\photoz{} performance is very good, with a scatter $\sigma_z < 0.015$ for SV-A1
data (Rykoff et~al., in prep).

\subsection{Photometric redshifts}
\label{sec:photo-z}

\begin{figure*}
\includegraphics[width=\linewidth]{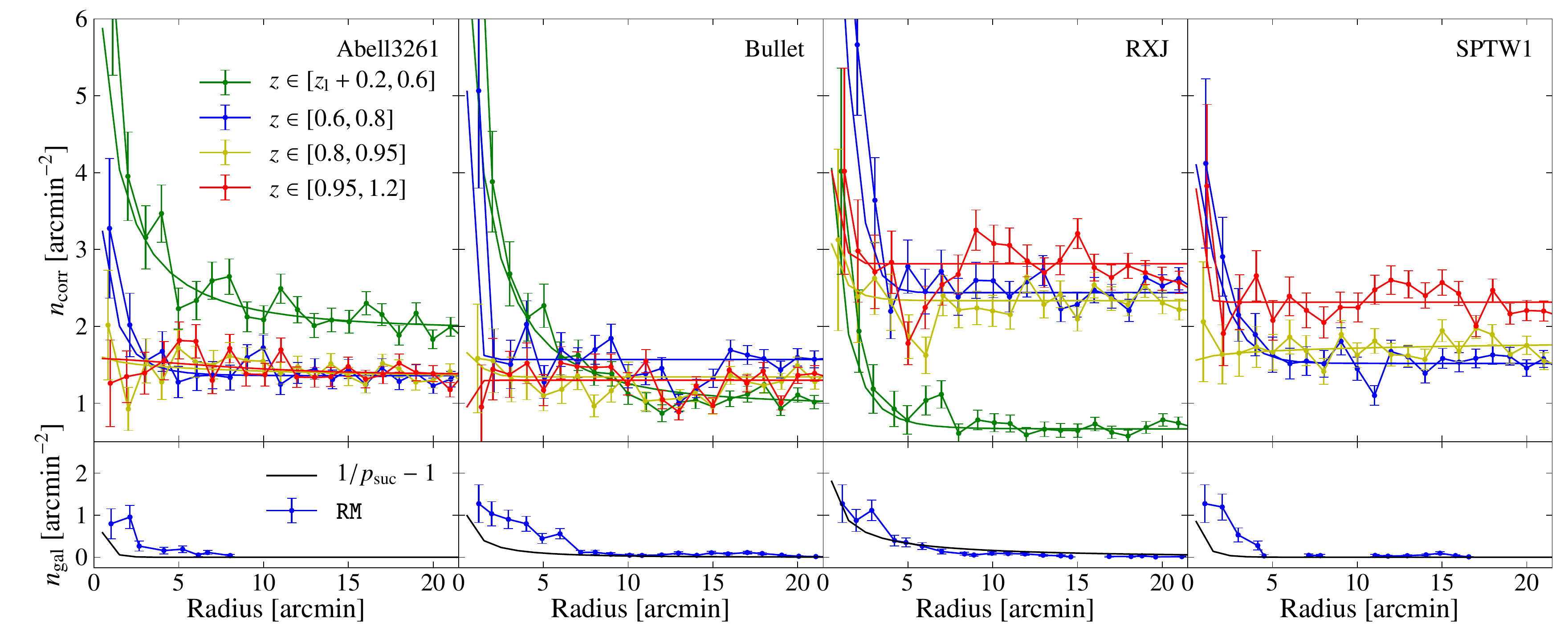}
\caption{\emph{Top:} Number density of galaxies with shape measurements (cf. \autoref{sec:shapes}) in several \photoz{} slices for each of the cluster fields. The solid lines are exponential fits to the corrected BG sample density (cf. \autoref{eq:cl_contamintation} and \autoref{fig:mag_distribution}). \emph{Bottom}: \redmapper-detected red-sequence galaxies in a narrow range around the cluster redshift (\emph{blue}), and the measured success-rate corrections $1/p_\mathrm{suc} - 1$  (\emph{black}), applied to the number densities shown above as per \autoref{eq:blending_corr}.}
\label{fig:number-profile}
\end{figure*}

We compute photometric redshifts (\photoz's) using the artificial neural network method that was applied to the Sloan Digital Sky Survey Data Release 6 (DR6) sample,
as described in detail by \citet{Oyaizu08.1}.  In brief, we use a neural network configuration with 10 input nodes, 
consisting of 5 $grizY$ \verb|MAG_AUTO| magnitudes and 
5 $grizY$ \verb|MAG_DETMODEL| magnitudes, followed by 3 hidden layers with
15 nodes per layer.  The neural network was trained using a set of about 12,000
galaxies with DES main-survey depth photometry and high-confidence 
spectroscopic redshifts.

The accompanying \photoz{} errors are computed using the empirical 
``nearest neighbor error'' (NNE) technique, described in detail
by \citet{Oyaizu08.2}.  The NNE method estimates the \photoz{} error for
each galaxy empirically, based on the \photoz's and redshifts of the 
galaxy's 100 nearest neighbors in the spectroscopic validation sample, 
where neighbor distance is defined using a simple flat metric in the space 
consisting of the 10 input magnitudes noted above.  Specifically, the 
NNE \photoz{} error $\Delta z_{\rm phot}$ is defined so that it corresponds to 68\% of 
the $|z_{\rm phot}-z_{\rm spec}|$ distribution of the nearest neighbors.  
In DES comparison tests by \citet{Sanchez14.1}, where the method we employ here is called {\sc desdm}, it shows a marginal bias $\langle\Delta z\rangle = -0.005 \pm 0.003$, a scatter of $\sigma_{68} = 0.094 \pm 0.002$, and a 3-$\sigma$ outlier rate of $0.018\pm 0.003$. All of these metrics indicate that the \photoz{} accuracy is easily sufficient for the purpose of this work. For further details, we refer the reader to \citet{Sanchez14.1}.

The \photoz s also serve us to estimate
\begin{equation}
\label{eq:sigmacrit}
\Sigma_{\rm crit} = \frac{c^2}{4\pi G}\frac{D_\mathrm{s}}{D_\mathrm{l} D_\mathrm{ls}},
\end{equation}
which will be used below to relate the gravitational shear
$\gamma$ to the surface density contrast of the cluster lens
\begin{equation}
\Delta\Sigma = \Sigma_{\rm crit}\ \gamma.
\end{equation}
In \autoref{eq:sigmacrit}, $D$ denotes angular diameter distance: to a source at photometric redshift $z_{\rm phot}$, to the lens at the spectroscopic redshift $z_\mathrm{l}$ (cf. \autoref{tab:clusters}; for \abell, we adopt the \redmapper{} estimate from \autoref{tab:results}), and between lens and source.

\subsection{Background galaxy selection}
\label{sec:bg_sel}

We select background galaxies according to their photometric redshifts $z_{\rm phot}$ by requiring that
\begin{equation}
\label{eq:bg_cut}
z_{\rm phot} > z_\mathrm{l} + 0.2.
\end{equation}
This seemingly straightforward selection is very effective at selecting a sample of galaxies that are behind the cluster, but due to the finite accuracy of the \photoz s it is not perfect. Two main limitations need to be addressed and are discussed below.

\subsubsection{Cluster-member contamination}
\label{sec:cl_contamination}

Even though we reject all \redmapper-detected galaxies in groups close to $z_\mathrm{l}$ from the background sample, we find that the number density of nominally background galaxies, for which we have both a \photoz{} and a shape estimate (see \autoref{sec:shape_comb} for details), rises strongly towards the cluster centers (cf. \autoref{fig:number-profile}). The rise is caused by cluster member galaxies, for which neither \redmapper{} nor the \photoz s are complete or accurate enough to put them at the cluster redshift. As expected, the chance of such galaxies getting upscattered to a particular redshift drops with increasing separation between $z_\mathrm{l}$ and $z_{\rm phot}$. 

Like \citet{Applegate12.1}, we correct for this effect by fitting an exponential model to the galactic number density as a function of cluster-centric distance $r$,
\begin{equation}
\label{eq:cl_contamintation}
n_{\rm corr}(r) = n_0 \left[1+\delta n_{C,Z} \exp\left[-\left(\frac{r}{r_{C,Z}}\right)^{\alpha_{C,Z}}\right]\right],
\end{equation}
where $\delta n_{C,Z}$, $r_{C,Z}$, and $\alpha_{C,Z}$ are free parameters and allowed to differ for each cluster $C$ and each \photoz{} slice $Z$\footnote{If any form of additional weighting is employed (as we will do with \autoref{eq:w3} later on),  $n_0$ needs to refer to the weighted number density.}. The slices are chosen to have a roughly constant and sufficient number of galaxies to allow a successful fit. The resulting fits are shown as solid lines in \autoref{fig:number-profile}. If we assume all of these contaminating galaxies to be randomly oriented,\footnote{Instrinsic alignments, in particular in combination with \photoz{} errors, could introduce non-random contributions to the lensing signal, but their strengths have been found to be insignificant for the work presented here \citep[e.g.][]{Chisari14.1,Sifon14.1}.} their effect is to reduce the perceived lensing effect of the background sample proportional to the contamination fraction. Following \citet{Blazek12.1}, we can therefore absorb the correction term into
\begin{equation}
\Sigma_{{\rm crit},C,Z} = \Sigma_{\rm crit} \left[1+\delta n_{C,Z}\exp\left[-\left(\frac{r}{r_{C,Z}}\right)^{\alpha_{C,Z}}\right]\right].
\end{equation}
Note that $\Sigma_{\rm crit}$ still depends on the actual (central) value $z_{\rm phot}$ of each source, but the correction term is necessarily averaged over \photoz{} slices of $\sim0.2$ width. Because the contamination fraction quickly rises when $z_{\rm phot}\rightarrow z_\mathrm{l}$, we expect that this best-effort correction is not entirely accurate, specifically for the lowest \photoz{} slices.

As detailed in \citet{Applegate12.1}, to obtain a meaningful $n_{\rm corr}$, one has to account for all effects that could change the number density, either related to the cluster or otherwise. First, masks around bright stars reduce the number of observable galaxies. At large distances, their impact is averaged out and only affects $n_0$. At smaller distance, we excised the radial range with prominent stars from the fit. This was necessary in the Bullet cluster field between 10 and 15 arcmin and in the RXJ field between 4 and 7 arcmin.

Second, the lensing-induced magnification could alter the observable number of galaxies, depending on the faint-end slope of the luminosity function. Even for clusters as massive as this, the magnification effect on the number counts is prominent only at small cluster-centric distances, which we will exclude from our lensing mass estimates in \autoref{sec:masses}, and is thus left uncorrected. 

Third, the high density of large galaxies in the core region of clusters prevents us from detecting background galaxies or measuring their shapes accurately. In other words, the success fraction of shape measurements declines towards the center, which could hide a substantial cluster-member contamination for the sample of galaxies we can measure shapes of. We assess the success probability $p_{\rm suc}(r)$ with the newly developed code \balrog\footnote{\url{https://github.com/emhuff/Balrog}} that allows us to insert artificial galaxies into our coadd images and compare the resulting catalogs with the input catalogs (Huff et~al., in prep). {In particular, we perform object detection and shape measurement, including any additional cuts, identically to the actual data analysis and then infer the rate of galaxies with acceptable shapes as a function of cluster-centric distance. Since we can only count galaxies \emph{after} their numbers $n_{\rm gal}$ have been reduced due to blending with the cluster galaxies, we need to correct for the effect according to
\begin{equation}
\label{eq:blending_corr}
n_{\rm corr}(r) = \frac{n_{\rm gal}(r)}{p_{\rm suc}(r)},
\end{equation}
which brings these numbers, and hence the parameters of the fit in \autoref{eq:cl_contamintation}, back to values they would have without the presence of cluster-member galaxies. The dimensionless term $1/p_{\rm suc} - 1$ is shown for each clusters as black line in the bottom panels of \autoref{fig:number-profile}, where we contrast it with the number density of \redmapper-detected galaxies (blue lines). We take the steeper profile of the $p_{\rm suc}$ boost factor as an indication that the main difficulty for measuring shapes in the central cluster regions stems from the existence of very large cluster galaxies, foremost the BCG, not just their increased number density.

We remark that the approach outlined here is very similar to the treatment in \citet{Applegate12.1}, in that the measured number densities are used to estimate the cluster-member contamination. But where they had to resort to proxies to estimate the success rate of their shape measurements, we actually \emph{measure} $p_{\rm suc}$ directly from our images.

\subsubsection{Photo-z inaccuracies}
\label{sec:photoz_calib}

The redshift estimate $z_{\rm phot}$ still enters directly into $\Sigma_{\rm crit}$, effectively treating $z_{\rm phot}$ as the true redshift of the source $z_s$. This is wrong in three ways. First, $\Sigma_{\rm crit}$ is a non-linear function of $z_s$, therefore even symmetric uncertainties $\Delta z_{\rm phot}$ as estimated in \autoref{sec:photo-z} lead to biased results for $\Sigma_{\rm crit}$ and thus for $\Delta\Sigma$, the net effect of which is an underestimate of $\Sigma_{\rm crit}$ that strongly rises with $z_{\rm phot} \rightarrow z_\mathrm{l}$. Second, one can furthermore imagine that occasionally $\Delta z_{\rm phot} > 0.2$, so that galaxies that are actually in front of the cluster make it into our background sample defined by \autoref{eq:bg_cut}. And third, the estimated \photoz s (or their errors) could be catastrophically wrong, so that we may misestimate $\Sigma_{\rm crit}$ with consequences yet to be determined.

We make use of the spectroscopic reference sample again and compute the true redshift distribution $p(Z_s\ |\ z_{\rm phot} \in Z)$ of sources in \photoz{} slices Z and spectroscopic slices $Z_s$ of width 0.1. Adopting a strategy closely related to \citet{Blazek12.1}, we determine the correction factor
\begin{equation}
c_Z^{-1} = \frac{\sum_{Z_s} p(Z_s\ |\ z_{\rm phot} \in Z)\ \Sigma^{-1}_{\rm crit}(Z_s)}{\Sigma^{-1}_{\rm crit}(Z)},
\end{equation}
whereby we mean $\Sigma_{\rm crit}(Z) = \bar{\Sigma}_{\rm crit}(z_{\rm phot}\in Z)$, and likewise for the spectroscopic slice $Z_s$. Applying this correction,
\begin{equation}
\label{eq:sigmacrit_final}
\Sigma_{{\rm crit}, C, Z, Z_s} = c_Z\ \Sigma_{{\rm crit}, C, Z},
\end{equation}
we control for the three \photoz{} errors on $\Delta\Sigma$ mentioned above at the level of $\Delta z_s=0.1$, which is more fine-grained than our cluster-member contamination correction from \autoref{sec:cl_contamination}. For the remainder of this work, we will only work with corrected $\Sigma_{\rm crit}$ values according to \autoref{eq:sigmacrit_final} without specifying it explicitly. A quantitative assessment of the amplitude of all corrections introduced in \autoref{sec:bg_sel} is given at the end of \autoref{sec:masses}.

\section{Weak-lensing analysis}
\label{sec:shape}

For this paper, we adopt a shape-measurement strategy, in which we perform the analysis on single-filter coadds and combine the results from the $r,i,z$ filters at the ellipticity level. While neither statistically nor systematically optimal,\footnote{The potential errors of the image coaddition could be avoided if the shape measurements are done on single-epoch images \citep[cf.][]{Miller13.1}, provided that the PSF can be modeled well with stars of lower significance. One can also reduce the noise-induced shape measurement biases by performing simultaneous shape fitting of all available exposures, even across filters. Both of these improvements are pursued for future lensing analyses in DES but go beyond the scope of this paper.} working with coadded images, commonly done in cluster lensing studies, allows us to perform the shape measurements -- of both stars and galaxies -- at relatively high significance, the importance of which will become evident in the next section.

\subsection{PSF modeling}
\label{sec:psfmodels}

\begin{figure*}
\includegraphics[width=\linewidth]{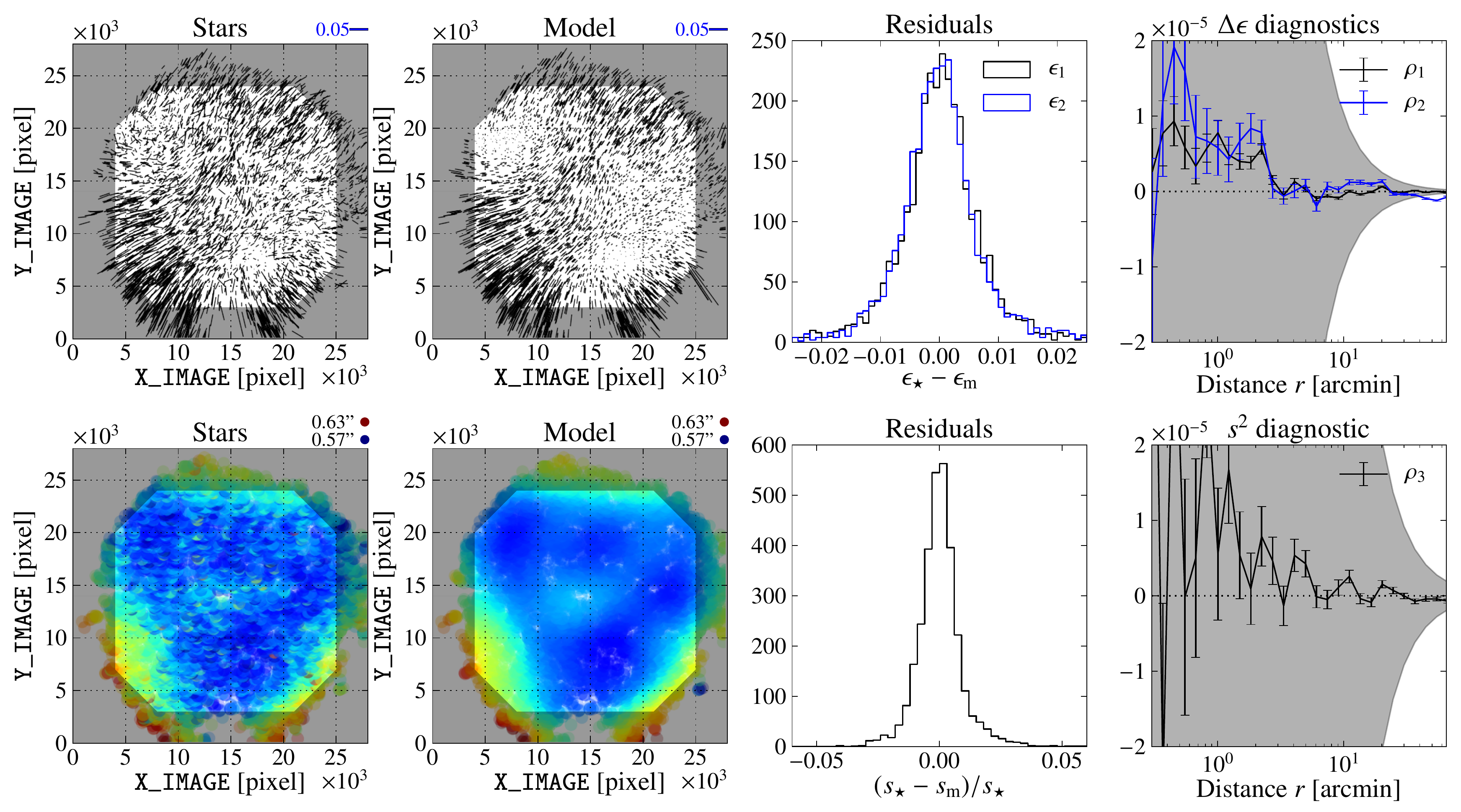}
\caption{Stellar and PSF model ellipticities and sizes for the \rxj{} $i.2$ coadd image. From \emph{left} to \emph{right}, the \emph{top} row shows the ellipticities $\epsilon_\star$ (in terms of shears, not polarizations, see e.g. Eq. 4.10 in \citet{Bartelmann01.1}) of the stars from the clean catalog (cf. \autoref{sec:starcatalog}); the corresponding ellipticities $\epsilon_\textrm{m}$ of the \psfex{} model derived from the same stars with \texttt{PSFVAR\_DEGREES=12}; the residuals between stellar and model ellipticities; the diagnostic two-point correlation functions $\rho_1$ and $\rho_2$ out to a maximum separation of 1 deg. The \emph{bottom} row is analogous but for stellar and PSF model sizes, respectively. See \autoref{sec:psftests} for details on the diagnostic functions and their tolerances (\autoref{eq:tolerances} and shaded areas in the right panels). In the left two panels, the shaded area indicates the field cut to eliminate parts of the coadd images not covered by all exposures. For reference, a comparison whisker and the size color scales are indicated in the top-right corner of these panels. The validation measurements are done with the moment-based shape code {\sc deimos} \citep{Melchior11.1}, i.e. we determine the moments of both stellar images and pixelized \psfex{} models at a fixed size of an adaptively matched elliptical Gaussian weight function of $\sigma_w=2.5$ pixels = $0\farcs658$.}
\label{fig:psfmodel}
\end{figure*}

Ordinarily, one models the PSF and its spatial variations simply by building a model from all available stars in the field. Unfortunately, PSF modeling for DECam is somewhat more complicated because its thick deep-depleted CCDs exhibit a mild flux dependence in the registration of charges. 
This is believed to be due to the accumulation of charges in the pixels altering the local electric field, effectively creating a repulsive force that scales linearly with the amount of charge already present \citep{Antilogus14}\footnote{We note that, in principle, the cosmic-ray rejection in our image coaddition procedure (cf. \autoref{sec:coadds}) can introduce a similar effect \citep[their section 2.2]{Heymans12.1}. However, the change of the stellar width as a function of stellar flux is consistent with the one observed by \citet{Antilogus14} on single-epoch images. We therefore conclude that our clipping procedure does not significantly affect the observed stellar widths.}.  The most apparent consequence is a flux dependence of the PSF width, hence the effect being dubbed the ``brighter-fatter relation''.  The effect is not quite isotropic, having a preferred alignment towards the readout direction.

A proper correction of this effect would involve modeling the redistribution of charges and locally re-assigning image counts between neighboring pixels to recreate the theoretical ``zero-flux'' shape of stars and galaxies. Such an approach is currently under development, but goes beyond the scope of this paper. In the following sections, we will adopt a simpler approach, in which we eliminate the brightest stars when building the PSF model (see \autoref{sec:brighter-wider} for details).  As they carry most of the photons, we need to compensate by pushing the star selection for the PSF model to fainter levels, where identification and shape measurement of stars can be performed much more reliably on coadded images. 

\subsubsection{Star selection and \psfex{} models}
\label{sec:starcatalog}
The first step of building a PSF model is to select a sample of stars, from which the shape of the PSF can be reliably inferred. Due to the large size of DECam, we need to be able to tolerate considerable variations of size and ellipticity of actual stars in this initial selection to avoid forming an incomplete model of the PSF. 

We work with the coadd catalogs from \sextractor{} and perform a first-pass selection of stars in the size-magnitude plane (to be precise, in the plane of \verb|MAG_AUTO| and both \verb|FWHM_IMAGE| and \verb|FLUX_RADIUS|), which yields mostly isolated stars, well suited for PSF measurements. To avoid saturated or noise-dominated stars, we could restrict the selection to objects with \verb|MAG_AUTO| $\in [15, 21.5]$, but the flux-dependence of the PSF forces us to introduce a much more restrictive selection \verb|MAG_AUTO| $\in [18, 21.5]$ to prevent the brightest stars from rendering the PSF model inappropriate for the bulk of fainter stars and galaxies (cf. \autoref{fig:brighter-wider} for an example with the full range of stellar magnitudes). 

We improve upon this first pass by requiring that stars be on the stellar branch in each of the filters $r,i,z$, which makes for a cleaner selection at faint levels and avoids the inclusion of blended stars whose faint companion is a drop-out galaxy for bluer filters. As a last step, we build a locally smoothed map of the \verb|FLUX_RADIUS| measurements of the stars selected so far and reject $3\sigma$ outliers. This localized selection is necessary for the wide-field imager DECam since stellar sizes increase considerably towards the edges of the field (cf. bottom-left panel of \autoref{fig:psfmodel}) so that mildly blended objects in the inner region could have passed the first selection with global size-magnitude cuts.

The procedure yields a very clean sample of objects, whose sizes are characteristic of relatively bright stars in the entire field with no noticeable contamination of neighboring objects. This selection is then passed on as input catalog to \psfex{} \citep{Bertin11.1}, run with \verb|BASIS_TYPE=PIXEL_AUTO|, \verb|BASIS_NUMBER=20| and varying polynomial degree \verb|PSFVAR_DEGREES| $\in \lbrace4,8,12,16\rbrace$. The pixel-based model is therefore formed by polynomial interpolation of $20\times20$ super-resolution cells, taken from $48\times48$ pixel cutouts.

\subsubsection{PSF modeling tests}
\label{sec:psftests}

\autoref{fig:psfmodel} shows one example of our PSF modeling approach, displaying stellar and model ellipticities in the top row and corresponding sizes in the bottom row. We can see that both sizes and ellipticities tend to increase towards the field edges and that there are structures present at various scales in both measurements. We therefore assess the validity of the \psfex{} models by the distribution of residuals and their cross-correlation functions, shown in the third and fourth column of \autoref{fig:psfmodel}, respectively.

Following \citet{Rowe10.1} (although using slightly different notation), we define the residual autocorrelation and the signal-residual cross-correlation functions of the (complex) ellipticity measurements,
\begin{equation}
\label{eq:D12}
\begin{split}
\rho_1(r) &\equiv \langle \conj{\Delta\epsilon_i} \Delta\epsilon_j\rangle_{i,j}\\
\rho_2(r) &\equiv \langle \conj{\epsilon_i} \Delta\epsilon_j+ \conj{\Delta\epsilon_i} \epsilon_j\rangle_{i,j},
\end{split}
\end{equation}
where the average comprises pairs of stars $i,j$ with separations $r$, and the residuals are defined as $\Delta\epsilon \equiv \epsilon_\star - \epsilon_m$, the difference of stellar and model ellipticities.  Conjugation of the complex ellipticity is notated as $\conj{\epsilon}$.

For the size measurements $s$,\footnote{by which we mean the sum of the flux-normalized second-order moments $s^2 = \frac{1}{F} (Q_{11} + Q_{22})$, or equivalently the intensity-weighted second moment of the radius $\langle r^2 \rangle_I$ averaged over image $I$.} we introduce a third diagnostic function similar to the above,
\begin{equation}
\label{eq:H1}
\rho_3(r) \equiv \Bigl\langle\Bigl(\frac{\Delta (s^2)}{s_\star^2}\Bigr)_i \Bigl(\frac{\Delta (s^2)}{s_\star^2}\Bigr)_j\Bigr\rangle_{i,j}
\end{equation}
based upon the fractional error in $s^2$. These two sets of diagnostic functions check for the anisotropic and the isotropic validity of the PSF model and hence for the amount of systematic shear misestimation introduced by insufficient PSF correction.

After defining the diagnostic functions, we need to answer the question: how small do they need to be? In \autoref{sec:diagnostics}, we work out the details, but the guiding principle is as follows. The error on the measured (deconvolved) shapes will be related to errors in the PSF model via a factor $T$ that compares the PSF size to the galaxy size. If we limit this PSF-induced shape measurement error by the intrinsic shape scatter of the background sample $\sigma_\epsilon$, which provides the fundamental limit to the statistical power of the lensing data, we can solve for the maximum tolerances of these diagnostics
\begin{equation}
\label{eq:tolerances}
\rho_1(r) + \left[\sigma_\epsilon^2 + \langle \conj{\epsilon_\star} \epsilon_\star\rangle(r)\right] \rho_3(r) < \frac{T^2\sigma_\epsilon^2}{n_{\rm gal}\ \pi (R_{\rm max}^2 - R_{\rm min}^2)},
\end{equation}
where $T=P_\gamma \bigl(\frac{s_{\rm gal}}{s_\star}\bigr)^2$, with $P_\gamma$ denoting the shear responsivity and $s_{\rm gal}$ the size of the galaxy \emph{prior} to convolution with the PSF, $n_{\rm gal}$ is the number density of galaxies with shape measurements, for which we adopt $n_{\rm gal}=10$ arcmin$^{-2}$ per filter as typical value (cf. \autoref{sec:shape_comb}).\footnote{In \autoref{sec:shape_comb} we will introduce additional lensing weights that effectively reduce  $n_{\rm gal}$ by 10-15\%, which means that our PSF diagnostics limits are slightly over-conservative.} The limits of the radial bin centered at $r$ are given by $R_{\rm max}$ and $R_{\rm min}$. As we point out in \autoref{sec:diagnostics}, we drop the requirement on $\rho_2$ as it does not exhibit substantial diagnostic power.

Measuring these diagnostics and comparing them to the maximum tolerance then allows us to determine how large a typical galaxy has to be so that the PSF-induced errors do not dominate over the shape scatter. In units of \shape's  \verb|FWHM_RATIO| (cf. \autoref{sec:shapes} for the definition and \autoref{eq:T} for the relation to $T$), our best models can use galaxies with \verb|FWHM_RATIO| $\geq1.1$. At our typical seeing of $\lesssim1$ arcsec, this limit corresponds to $s_{\rm gal}\approx0.4$ arcsec, in line with measurements for the typical sizes of galaxies at our depth based on deep HST imaging in \citet[cf. their Fig. 1]{Miller13.1}. We do not consider PSF models, where that requirement would go beyond \verb|FWHM_RATIO| $=1.2$ (or $s_{\rm gal}>0.65$ arcsec), the size cut, for which the shape measurement code is well tested  (see next section). All fields and filters that qualify under this requirement are listed with a $\checkmark$ in \autoref{tab:seeing}. 

In terms of complexity, the best-performing \psfex{} models have
polynomial degree between 8 and 16, with a majority at 16. In addition
and as replacement of the proposed role of $\rho_2$ in
\citet{Rowe10.1}, we performed a cross-validation study, in which we
build a model using only a subset of the stars and then compute the
diagnostic functions from the remaining stars. Even with the highest
polynomial degrees, the models showed no indication of over-fitting
the data and, the with the numbers of stars we provided, produced stable results in the analyzed areas of each field.

To our knowledge, this is the first time that tolerances for PSF model diagnostics have been utilized to predict what size galaxies need exceed so that their shapes can be sensibly determined. The approach we have adopted here is conservative in three distinct ways. First, because we require the PSF models to stay within their tolerances at all scales, the actual PSF errors are smaller than our limits \emph{at most scales}. Second, by calculating the tolerance at the limit of the smallest galaxies, the majority of the source sample will be less affected by PSF errors than predicted. And third, as we show in \autoref{sec:shape_comb}, the final shape catalogs are combined from different filters, so that the variation of PSF properties between filters enables a partial cancellation of the PSF-induced errors. We therefore consider the resulting shapes not to be dominated by PSF systematics, a claim we are going to review in \autoref{sec:shear_sys}.

\subsection{\shape{} measurements}
\label{sec:shapes}

For the weak-lensing shear analysis presented in this
paper, we use the publicly available galaxy shape measurement code
\shape\footnote{\url{https://bitbucket.org/joezuntz/im3shape/}, revision  \href{https://bitbucket.org/joezuntz/im3shape/commits/c8e6728335b16c8b2acdab43a073340069813e69?at=SVCluster_on_legion}{c8e6728}} \citep{Zuntz13.9}. By maximizing the likelihood, it fits
a PSF-convolved two-component bulge-plus-disc galaxy model to measure
the ellipticity of each galaxy. In particular, we model galaxies as a
sum of co-elliptical \citet{Sersic63.1} profiles described by seven free parameters:
ellipticity ($\epsilon_1$, $\epsilon_2$), position ($x_0$, $y_0$),
disc half-light radius ($r_\mathrm{d}$), bulge and disc
  peak flux ($A_\mathrm{b}$, $A_\mathrm{d}$). We set the indices of the \sersic{}
profiles to 1 for the disc component and 4 for the bulge
component. The bulge-disc size ratio is also kept fixed at 1.0.

To counter the adverse effects of aliasing and avoid upsampling biases
we render intermediate model images at higher resolution as described
in detail in 4.1 of \citet{Zuntz13.9}, choosing \shape{}'s upsampling
parameters conservatively as \verb|upsampling=5|,
\verb|n_pixels_to_upsample=8|, and
\verb|n_central_pixel_upsampling=7| with a postage stamp of size \verb|stamp_size=37|
pixels. This is done for both the galaxy model image and the PSF image
sampled from the PSFEx model at the galaxy position estimated with
\sextractor. The convolution of the two is then performed in Fourier space.

There are notable complications in our use of \shape{}
here compared to the simulation study presented by \citet{Zuntz13.9}:
{
\renewcommand{\descriptionlabel}[1]{\hspace{\labelsep}\textit{#1}.}
\begin{description}
\item[Neighboring objects] The flux of nearby objects affects shape measurement and leads to biases if not treated carefully. To this end, we make use of the segmentation map provided by \sextractor: we give zero weight to all pixels that are assigned to another identified object within the processed image stamp.
\item[Background treatment] Although the single-filter coadd
  images are globally background-subtracted by \swarp{}, we find a
  non-zero local background for a few postage stamps where the global
  background subtraction remained insufficient. Therefore, we perform a local background estimation by averaging those pixels within the postage stamp that have not been assigned to any detection. The resulting value is then subtracted as a constant from all pixels within the
 analyzed image.
\end{description}
}

We process each single-filter coadd image independently. For each
image we run on all detected objects (with the exception of very
bright and very faint objects). To clean the final shear catalogs, we first perform a star-galaxy separation
based on the \shape{}-estimated \verb|FWHM_RATIO| between the
pixel-PSF-convolved model image and the PSF image. The FWHM
of a pixel-PSF-convolved galaxy model is estimated from its centered,
noise-free model image with its ellipticity set to zero. The \verb|FWHM_RATIO| is in effect a measure of the pre-seeing mean radius of the source, and hence we expect cuts based on this quantity to be free of first-order biases toward alignment with the PSF. All
objects with \verb|FWHM_RATIO| $< 1.1$ are considered
unreliable for shape measurement, often stars, and are excluded from further processing. 
We also remove shape measurement outliers by applying
additional cuts based on \shape{} fit results. In particular,
we apply cuts for the following parameters: the best-fit likelihood value, the minimum
and maximum model value (per pixel), the minimum and maximum residual value (per pixel), the change in the estimated centroid position, and the disc half-light radius. The cuts are adjusted for each coadd image and are only meant to reject obvious failures of the shape measurement process.

\subsubsection{Noise-bias calibration}
\label{sec:noise-bias}

Shape measurements are affected by a prominent bias when the galaxy images become noisy \citep[e.g.][]{Massey07.1}. This is a consequence of the observable, the galaxy ellipticity, being non-linearly related to the flux in each pixel and applies to model-fitting methods and moment-based measures of the ellipticity alike \citep{Melchior12.1, Refregier12.1, Kacprzak12.1}.

To calibrate \shape's response to noise bias we simulate mock galaxies, using the {\sc GalSim}\footnote{\url{https://github.com/GalSim-developers/GalSim}} \citep{Rowe14.1} framework. In particular, we adopt the methodology of \citet{Mandelbaum12.1} and degrade high-resolution and high-significance images from COSMOS to the DECam resolution and magnitude limit (cf. \autoref{fig:mag_distribution}). We approximate the coadd PSF by a circular \citet{Moffat69.1} profile with seeing values $\in [0.7, 0.8, 0.9]$ arcsec, spanning the expected range of our observing conditions. Applying exactly the same cuts as for shape catalogs from the coadd images, we have verified that both magnitude and size distributions of the simulated galaxies closely match the observed ones.
Adding an artificial shear $\gamma$ of order 5\%, we can infer the shear response 
\begin{equation}
\label{eq:shear_response}
m_{\rm n} \equiv \frac{\partial \langle\epsilon\rangle}{\partial\gamma}
\end{equation}
as a function of the signal-to-noise ratio \verb|SNR| and \verb|FWHM_RATIO|. We believe these two parameters to largely determine the shear response, and by working with size ratios we render this calibration mostly insensitive to the observed PSF widths that occasionally exceeded the simulated range. The result is shown in \autoref{fig:noise-bias}. At high \verb|SNR|, the shear can be measured in an unbiased fashion for all galaxy sizes, whereas the noise bias gets progressively worse for lower \verb|SNR|, scaling roughly as \verb|SNR|$^{-2}$, consistent with findings of \citet{Bernstein02.1}. It is counter-intuitive that the smallest galaxies show the least amount of bias. Also, at very low \verb|SNR| the larger galaxies show an intriguing upturn.  We interpret both as higher order effects of the noise bias. A decrease in noise bias for smaller galaxies is plausible, particularly for a mixture of biases with different signs at different orders in \verb|SNR| and \verb|FWHM_RATIO|, with perhaps some fortuitous cancellation for the smallest galaxies we plot. According to \citet{Kacprzak12.1}, only even orders of \verb|SNR| can appear in the noise-bias relation, therefore we attempt to parameterize the dependence with the following polynomial model,
\begin{equation}
\label{eq:noise-bias-fit}
m_\textrm{n} \approx c_0 + c_2 \texttt{SNR}^{-2} + c_4 \texttt{SNR}^{-4},
\end{equation}
whose best-fit parameters are listed in \autoref{tab:noise-bias-fit}. The applied noise-bias corrections are taken from these fits in each of the size bins.

\begin{figure}
\includegraphics[width=\linewidth]{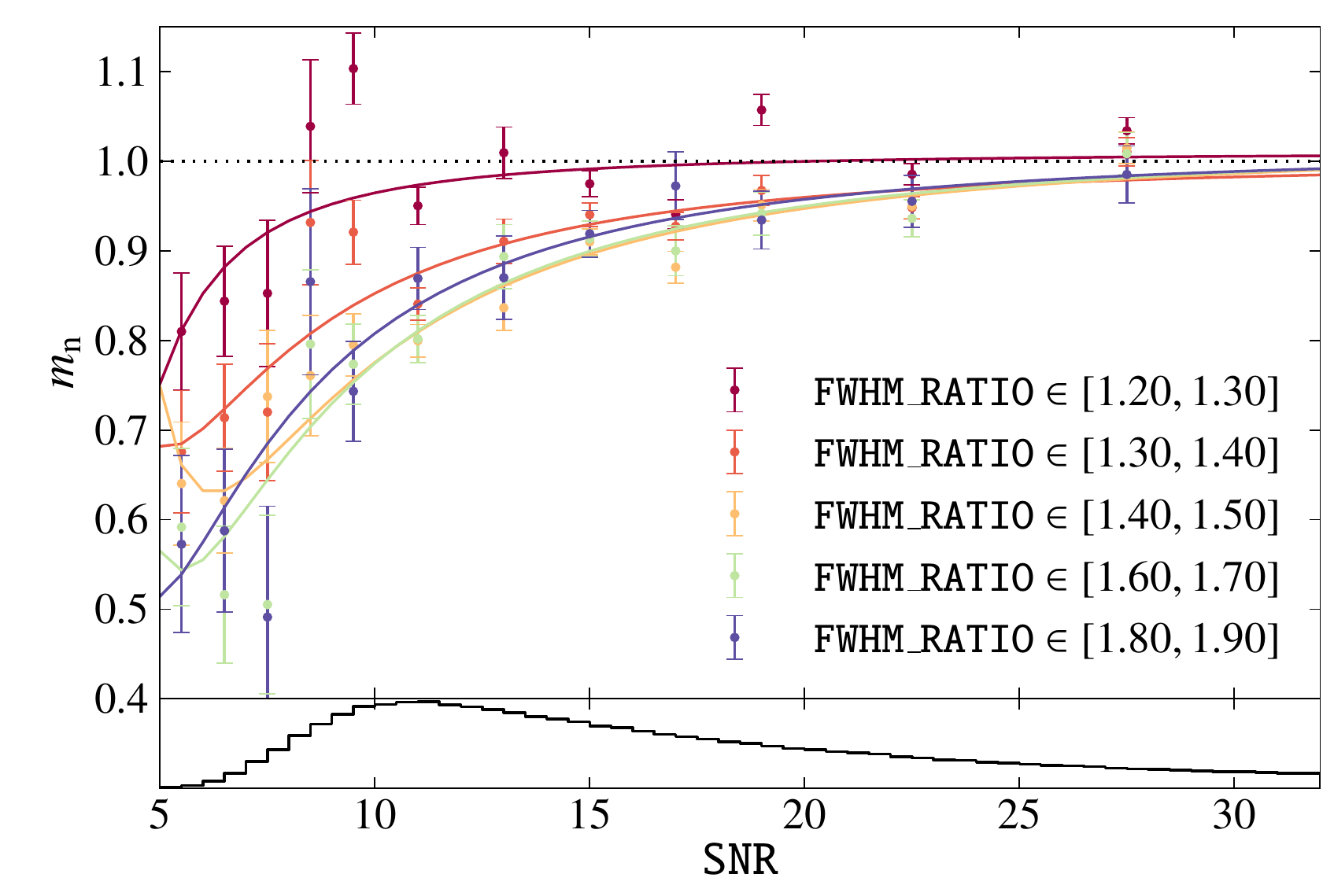}
\caption{Noise bias on the multiplicative term $m_\textrm{n}$ of the shear response as a function of \shape's $\texttt{SNR}$ for different values of the galaxy \texttt{FWHM\_RATIO}. The solid lines are even-order polynomial fits to the data (cf. \autoref{eq:noise-bias-fit}). The bottom panel shows the \texttt{SNR} distribution of galaxies with shape measurement from any single-filter coadd image, averaged over all fields and $riz$ filters.}
\label{fig:noise-bias}
\end{figure}

We also note that the shear response of \autoref{eq:shear_response} is assumed to be linear in the shear. Large shears, as encountered close to the centers of clusters, may introduce a non-linear response, which would be overseen by our calibration approach. However, for the analysis in \autoref{sec:profile} we will exclude these inner regions and restrict ourselves to shear values, for which \shape's performance was tested by our simulations and others \citep[e.g.][]{Kitching12.1}.

\subsection{Final catalog creation}
\label{sec:shape_comb}

\begin{figure*}
\includegraphics[width=\linewidth]{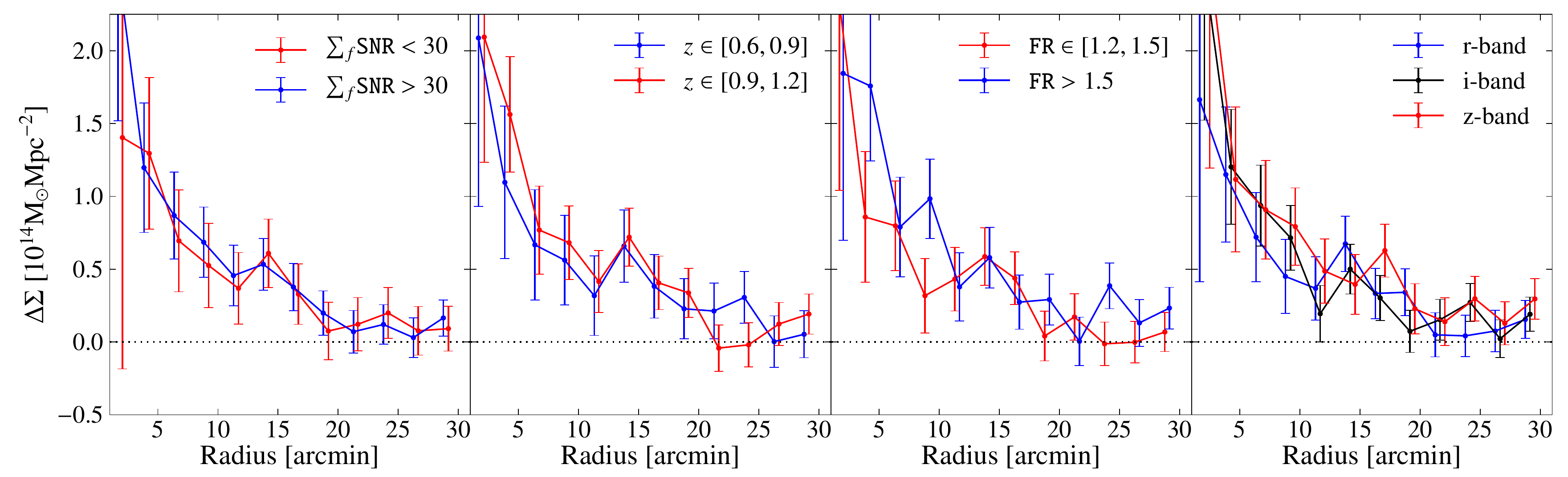}
\caption{Consistency test for the background-selected shape catalogs, sliced in \shape{}'s $\sum_f$\texttt{SNR} (\emph{1st panel}), in \photoz{} estimate (\emph{2nd panel}), in \texttt{FWHM\_RATIO} (\texttt{FR} for short, \emph{3rd panel}), and from different filters (\emph{4th panel}). The conversion between shear estimate and $\Delta\Sigma$ is done with $\Sigma_{\rm crit}(z)$ calculated for each galaxy individually based on its photometric redshift according to \autoref{eq:sigmacrit_final}. The numbers are stacked over the four fields, errors correspond to the dispersion of the weighted mean in each bin. For clarity, the points of different slices have been shifted horizontally by 0.4 arcmin.}
\label{fig:consistency}
\end{figure*}

For each cluster field we include the shape catalogs for all filters $f$ that passed the PSF modeling requirements from \autoref{sec:psftests} (cf. \autoref{tab:seeing} for the list). Based on \sextractor{} measurements, we enforce additional cuts to clean the catalogs from potentially problematic measurements: \verb|FLAGS| $=0$ and \verb|CLASS_STAR| $<0.8$. We furthermore exclude areas at the edges of the fields, where the coverage is not homogeneous across filters, giving rise to the ``picture frame'' geometry shown in the left panels of \autoref{fig:psfmodel}. We also mask out stars detected in the Tycho-2 catalog \citep{Hog00.1} with magnitude-dependent circular and hand-crafted bleed trail masks  to avoid saturation features and local sky-background variations that could affect the photometry or the shape measurements. The remaining sample is  shown as ``Clean" in \autoref{fig:mag_distribution}.

\begin{table}
\caption{Best-fit parameters of \autoref{eq:noise-bias-fit} to the simulated data from \autoref{fig:noise-bias}. The last column indicates the percentage of all galaxies with shape measurements in any band to fall into the given bin, averaged over all fields and $riz$ filters. The missing fraction of about 7\% are larger than those listed here but show an identical noise bias, we therefore adopt the $\texttt{FWHM\_RATIO} = 1.9$ fit to correct them.}
\label{tab:noise-bias-fit}
\begin{tabularx}{\linewidth}{lllll}
\verb|FWHM_RATIO| & $c_0$ & $c_2$ & $c_4$ & $N_\mathrm{gal}$ [\%]\\
\hline
$\in[1.20,1.30]$ & $1.010\pm0.012$ & $-3.9\pm3.6$ & $-64\pm129$ & 24.2\\
$\in[1.30,1.40]$ & $1.001\pm0.012$ & $-17.2\pm3.4$ & $229\pm127$ & 19.0\\
$\in[1.40,1.50]$ & $1.019\pm0.013$ & $-30.2\pm3.5$ & $587\pm130$ & 15.3\\
$\in[1.50,1.60]$ & $1.028\pm0.016$ & $-30.5\pm4.1$ & $498\pm150$ & 12.0\\
$\in[1.60,1.70]$ & $1.019\pm0.018$ & $-28.9\pm4.8$ & $438\pm174$ & 9.8\\
$\in[1.70,1.80]$ & $1.042\pm0.021$ & $-28.3\pm5.4$ & $326\pm193$ & 7.3\\
$\in[1.80,1.90]$ & $1.014\pm0.024$ & $-23.4\pm6.2$ & $271\pm214$ & 5.0\\
\hline
\end{tabularx}
\end{table}

On top of the selection inherent to \shape{} success, we restrict the shape measurements to \verb|SNR| $>5$ and \verb|FWHM_RATIO| $>1.2$ in concordance with limits on both the \psfex-model quality and the shape calibrations in \autoref{sec:noise-bias}.
For each remaining galaxy $j$, the shapes are combined from all available filters $f(j)$ according to their \shape{} weight $w_3 (j, f)$,
\begin{equation}
\label{eq:shapes_combined}
\bepsilon (j) = \frac{\sum_{f(j)} m_\mathrm{n}(j, f)^{-1}\ \bepsilon (j, f)\ w_3 (j, f)}{\sum_{f(j)} w_3(j, f)}.
\end{equation}
This filter-combined ellipticity maximizes the number of galaxies with shape measurements and reduces the variance from pixel noise in each ellipticity estimate derived from more than a single filter. The resulting catalog forms the basis for all subsequent analysis and is denoted as the ``Shapes" sample in \autoref{fig:mag_distribution}. The number density $n_\textrm{gal}$ of this catalog ranges from 9 to 12 arcmin$^{-2}$, with the best available seeing in each field being the dominant factor of that variation. This is consistent with the expectations for the full-depth DES imaging data \citep{DES05.1}. Only the coadd images of the additional cluster \abell{} do not reach the nominal full depth of 10 exposures, resulting in a reduced $n_\textrm{gal}=8$ arcmin$^{-2}$.

Note that the noise-bias correction $m_n(j, f)^{-1}$ in \autoref{eq:shapes_combined} depends on $\texttt{SNR}(j, f)$ and $\texttt{FWHM\_RATIO}(j, f)$ as described in \autoref{sec:noise-bias}. The weight takes both statistical and measurement variances into account \citep[e.g.][their Eq. A2]{Hoekstra00.1},
\begin{equation}
\label{eq:w3}
w_3(j, f) = \frac{\sigma_\epsilon^2}{\sigma_\epsilon^2 + \Bigl[0.1\frac{1}{m_\mathrm{n}(j, f)}\frac{20}{\texttt{SNR}(j, f)}\Bigr]^2},
\end{equation}
where we have adopted an estimator for the measurement error $\sigma_j$ that scales inversely with \verb|SNR| and accounts for variable amounts of noise-bias correction. This estimate may be on the conservative side, but has performed well in simulations, where a full likelihood exploration for the parameters was available. For the weight $w_3(j)$ of the combined shapes, we stick to the formula above, but replace $\texttt{SNR}(j, f)$ with $\sum_{f(j)} \texttt{SNR}(j, f)$.

When we apply the weights to the shape catalog, the effective number density $n_\mathrm{eff}\approx 0.87\, n_\mathrm{gal}$, while both shape scatter and mean redshift of all galaxies with shape measurements are only mildly (i.e. of order 5\%) reduced by the weighting, $\langle z_\mathrm{phot} \rangle \approx 0.7$ and $\sigma_\mathrm{e} \approx 0.3$. We emphasize that because of anticipated changes in the shear-measurement pipeline for future studies, these numbers are only roughly indicative of the DES performance and should therefore be treated with due caution.

Applying the \photoz{} cut of \autoref{eq:bg_cut} finally yields the shape catalog for the background sample, labelled as ``BG" in \autoref{fig:mag_distribution}.

\vspace{-10pt}
\subsection{Consistency tests}
\label{sec:shear_sys}

While not strictly necessary for the lensing analysis in \autoref{sec:profile}, we choose to express the measurements in $\Delta\Sigma(r)$ rather than in the actual observable, the shear $g = \langle \epsilon\rangle$. This physical quantity should be invariant under choice of source populations, at least ideally. It allows us to slice the background sample in various ways and thereby to test whether the shear measurements and the various calibrations are accurate. To work with a sufficient number of galaxies, we combine the four clusters by stacking them at their respective (BCG) centers. The test results are shown in \autoref{fig:consistency}. 

First, we check whether the typical SNR-dependence of weak-lensing measurements is corrected by our calibrations from \autoref{sec:noise-bias}. The left panel of \autoref{fig:consistency} shows that there is no strong trend visible when varying \shape{}'s \verb|SNR| parameter, implying that our calibration was indeed successful. In terms of weak-lensing mass (see methology in \autoref{sec:masses}), we have less than 10\% variations between each subset and the entire stack.
Note that we slice the final catalog, where ellipticity measurements have been combined according to \autoref{eq:shapes_combined}, therefore the SNR is given as the sum over the measurements in each filter. The pivotal value $\sum_f\texttt{SNR}=30$ is chosen here because it yields $\texttt{SNR}(f)\approx 10$, which, according to \autoref{fig:noise-bias}, separates galaxies with substantial levels of noise-bias corrections from those with a much milder correction.

Second, we split the sample according to the reported \photoz{} of each source. As we have corrected for the redshift dependence of the measured shear by converting to $\Delta\Sigma$ (see \autoref{sec:bg_sel} for the details and calibrations we applied to the raw \photoz{} values), we should not see any variation induced by the change in distances in \autoref{eq:sigmacrit}, and indeed there is none recognizable in the second panel of \autoref{fig:consistency}. Quantitatively, we find the upper redshift slice to have a higher mass estimate of about 15\%, whereas the lower is about 10\% low in mass compared to the entire stack, neither of which is significant given the errors in the stacked lensing profiles. Note that this outcome is not trivial as both the \photoz{} corrections and the noise-bias calibrations have to perform well. Due to the correlation between flux and distance, correcting only one of them is not sufficient to null this test.

Third, we revisit our claim from \autoref{sec:psftests} that the \psfex{} models allow us to use galaxies down to \verb|FWHM_RATIO| $=1.2$ or even below. The third panel of \autoref{fig:consistency} shows that this is unfortunately not entirely the case, with a mild size dependence of the reported $\Delta\Sigma$: the larger (smaller) set of galaxies yields a mass estimate that is about 30\% higher (10\% lower) than the entire stack. While at rather low statistical significance -- the errors on the mass estimate of each slice are of order 35\% -- we suspect that correlated noise and  complex PSF shapes in the coadd images are more harmful to small galaxies than indicated by the noise-bias simulations from \autoref{sec:noise-bias} that used uncorrelated pixel noise and simplistic PSFs. We want to point out that this tendency is barely recognizable in a stack of four clusters, so that we do not expect it to limit the individual lensing analyses in the next sections.

Finally, the last panel of \autoref{fig:consistency} shows the lensing signal if we only use the shape measurements from single filters instead of combining them according to \autoref{eq:shapes_combined}. Using single filters constitutes a drop-out technique, where galaxies are more likely measurable in redder filters if they are at higher redshifts. Since the redshift dependence of the signal seems to be well characterized (second panel), we expect consistent measurements here, too. However, uncorrected effects related to the CCD (e.g. prominent fringing in the $z$-band) or the instrument in general could interfere, but to the limit of this test we can rule this out: the mass estimates agree to better than 10\% across filters. This leads us to the non-trivial conclusion that DECam images taken in each of the $riz$ filters seem equally suitable for shape measurements. 

Note that this methodology effectively constitutes a sequence of null tests, even though we inspect the actual signal. We could have subtracted the mean signal to render it a proper null test, but we choose to leave it in since some of the systematics could scale with, for example,  the lensing strength or the source density, so it may help to actually see the mean cluster signal to gauge the dependency on cluster-centric distance $r$. Larger cluster samples investigated in forthcoming DES analyses will substantially sharpen these consistency tests and allow us to detect potential shape measurement problems with much higher precision.

\section{Shear profiles and lensing masses}
\label{sec:profile}

\begin{figure*}
\includegraphics[width=\linewidth]{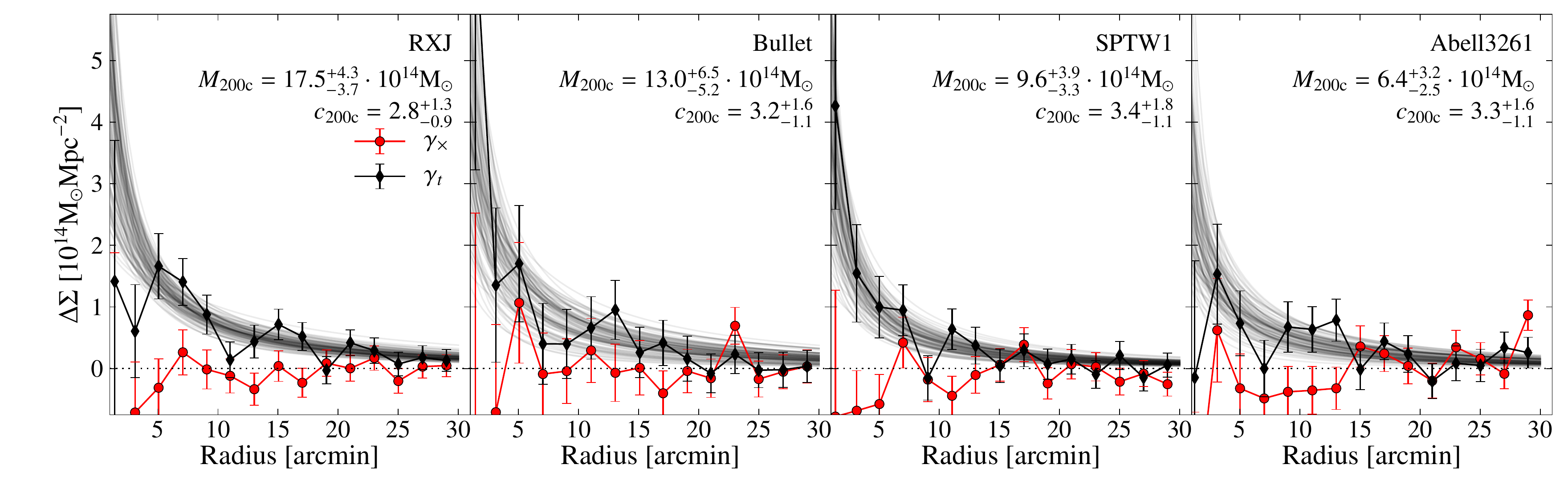}
\caption{Surface density contrast $\Delta\Sigma=\Sigma_\textrm{crit}\gamma_\mathrm{t}$ profiles (\emph{black}) for each of the four cluster fields and 100 random MCMC sample projections onto the data (\emph{light gray}) after an initial burn-in phase. The fit range was restricted to $3\leq r\leq 15$ arcmin. The B-mode $\Sigma_\textrm{crit}\gamma_\times$ is shown in \emph{red}.}
\label{fig:shear_profiles}
\end{figure*}

We now measure the lensing masses by fitting a radial profile to the tangential shear signal, centered on the BCG coordinates as listed in \autoref{tab:clusters}. 

\subsection{NFW profile fits and lensing masses}
\label{sec:masses}

To obtain estimates of cluster masses, we assume the density profile described first by \citet[][NFW]{Navarro96.1}. The three-dimensional density $\rho(r)$ of the NFW profile at radius $r$ is given as
\begin{equation}
\rho(r)=\frac{\rho_0}{(r/r_\mathrm{s})(1+r/r_\mathrm{s})^2} \; .
\label{eqn:nfwrho}
\end{equation}
The profile can alternatively be expressed in terms of the mass $M_{200\mathrm{c}}$ and concentration $c_{200\mathrm{c}}=r_{200\mathrm{c}}/r_\mathrm{s}$, instead of the central density $\rho_0$ and scale radius $r_\mathrm{s}$. Here $r_{200\mathrm{c}}$ denotes the radius of a sphere that comprises an overdensity of 200 times the critical density at the redshift of the cluster. The projected density and gravitational shear of the NFW profile are given in \citet{1996AuA...313..697B} and \citet{2000ApJ...534...34W}.

Assuming Gaussian errors on the shape estimates,\footnote{That is an overly simplified assumption because measurement errors will induce Cauchy-like wings even if the intrinsic shape dispersion were Gaussian \citep{Applegate12.1, Melchior12.1}.} the likelihood of any model can be calculated from the shear catalogue by means of the $\chi^2$ statistic. Given model predictions $\Delta\hat\Sigma (r)$ for the lens and the measurements of component $\epsilon_\mathrm{t}(j)$ of the ellipticity of galaxy $j$ tangential to the cluster center, the likelihood $\mathcal{L}$ can be written as
\begin{equation}
\ln \mathcal{L}=-\frac{1}{2}\sum_j\frac{\bigl[\Delta\hat{\Sigma} (r_k) -\Sigma_{\rm crit}(j)\epsilon_\mathrm{t}(j)\bigr]^2}{\Sigma_{\rm crit}^2(j)\bigl[\sigma^2_j+\sigma^2_\epsilon\bigr]} + \mathrm{const},
\label{eqn:likelihood}
\end{equation}	
where we use the corrected $\Sigma_{\rm crit}$ from \autoref{eq:sigmacrit_final} and insert $w_3(j)/\sigma_\epsilon^2$ from \autoref{eq:w3} as the total variance term in the denominator. Since the reduced shear in the weak-lensing regime is a small correction to the intrinsic shape, the latter can still employ $\sigma_\epsilon$ from the observed (as opposed to an unlensed) distribution of shapes. We evaluate the posterior distribution of the likelihood $\mathcal{L}$
with the MCMC sampler {\sc
  emcee}\footnote{\url{http://dan.iel.fm/emcee/}}
\citep{Foreman-Mackey13.1}, adopting a log-normal prior on the concentration following the
concentration-mass relation of \citet[their 'full' sample]{2008MNRAS.390L..64D} with scatter of $\sigma_{\log c}=0.18$ \citep{2001MNRAS.321..559B}. 

To avoid the regions where the cluster-member contamination correction (cf. \autoref{sec:cl_contamination}) and possible shape-measurement errors due to crowding may not be well characterized, we exclude the region in the very center and start the fit at $r=3$ arcmin. This also renders us robust against miscentering as our choice of the BCG center may not correspond to the actual gravitational center of the halo. To limit the inclusion of uncorrelated large-scale structure or clusters, we also restrict the outer limit to $r=15$ arcmin. The resulting range $3\leq r \leq 15$ arcmin is similar to $750\ \mathrm{kpc} \leq r \leq 3\ \mathrm{Mpc}$ employed by \citet{Applegate12.1} but extends somewhat farther out to reduce the shot noise from the rather low $n_{\rm gal}$ of our data. The NFW profile is not a good fit to lensing measurements at such large distances because it lacks the two-halo contribution from structures associated with the clusters. However, the resulting bias is only of order 10\% \citep[their Fig. 4]{Oguri11.1}, which will certainly be below our measurement accuracy. 

In \autoref{fig:shear_profiles} we show the individual shear profiles and 100 randomly chosen sample projections onto the data to demonstrate the range of viable models after an initial burn-in phase. Parameter confidences are given in terms of the 16th, 50th, and 84th percentiles of the marginalized mass $M_{200\mathrm{c}}$ and concentration $c_{200\mathrm{c}}$ distributions. We can see that SPTW1 is well-fit by an NFW model, including the innermost radial bin that was not included in the fit. Given our uncertainties, the NFW profile constitutes an acceptable model for all clusters. We also want to point out that the B-mode, denoted as $\gamma_\times$ in \autoref{fig:shear_profiles}, is statistically consistent with zero for all clusters, although some moderately large fluctuations occur.

\begin{table*}
\caption{Weak lensing masses $M_{200\textrm{c}}$ in units of  $10^{14} \mathrm{M}_\odot$ (with a log-normal prior on $c_{200\mathrm{c}}$ based on the \citet{2008MNRAS.390L..64D} concentration-mass relation), \redmapper{} richness $\lambda$ and redshift estimate $z_{\lambda}$, and their statistical errors (see \autoref{sec:cl_sel} and \autoref{sec:masses} for details). The literature mass estimates are derived from weak lensing, galaxy dynamics (D) or optical richness (R).}
\label{tab:results}
\begin{autopn}
\renewcommand{\arraystretch}{1.3}
\setlength{\tabcolsep}{.4em}
\begin{tabular}{lllll}
Cluster name  & $M_{200\mathrm{c}}$ & $\lambda$ & $z_{\lambda}$ & Literature value $M_{200\mathrm{c}}$\\
\hline
\rxj & $17.5^{+4.3}_{-3.7}$ & $203 \pm 5$ & $0.346 \pm 0.004$ & $22.8^{+6.6}_{-4.7}$ \citep{Gruen13.1}, $20.3 \pm 6.7$ \citep{Umetsu14.1}, $16.6\pm1.7$ \citep{Merten14.1}\\
\bulletcl & $13.0^{+6.5}_{-5.2}$ & $277 \pm 6$ & $0.304 \pm 0.004$ & 17.5 \citep{Clowe04.1}\parnote{We converted the measured $r_{200\mathrm{c}}$ from \citet{Clowe04.1}, which lacks an error estimate, to $M_{200\mathrm{c}}$ using the critical density in our adopted cosmology.}, $12.4$ \citep[D]{Barrena02.1}\\
\sptw & $9.6^{+3.9}_{-3.3}$ & $77 \pm 4$ & $0.391 \pm 0.008$ & $11.2^{+3.0}_{-2.7}$ \citep{Gruen13.2}, $4.9\pm3.3\pm1.4$ \citep[R]{High10.1}\\
\abell & $6.4^{+3.2}_{-2.5}$ & $71 \pm 3$ & $0.216 \pm 0.003$ & ---\\
\hline
\end{tabular}
\end{autopn}
\end{table*}

It is typical for pure weak-lensing measurements that the concentration is only poorly constrained \citep[e.g.][]{Postman12.1}, a tendency that we have even exacerbated by excluding the inner 3 arcmin. This highlights the potential importance of a concentration prior, a situation in which the significant differences in the literature between concentration-mass relations derived from different simulations or observational studies may appear worrisome.
However, deviations of the assumed relation
from the truth only mildly impact the weak-lensing mass measurement
\citep[e.g.][their section 4.3]{2012MNRAS.427.1298H}. Indeed, we find
no significant differences of the marginalized results using the Duffy
prior or an entirely flat prior within $0< c_{200\mathrm{c}}< 8$.

Comparing our $M_{200\mathrm{c}}$ estimates with previous results listed in \autoref{tab:results}, often based on substantially deeper data, we find good agreement for \rxj{}, where the mass estimate in \citet{Gruen13.1} is within our 68\% confidence region. Two recent analyses of the same data -- together with magnification \citep{Umetsu14.1} or \emph{HST} strong- and weak-lensing constraints \citep{Merten14.1} -- yield reduced estimates of $M_{200\mathrm{c}}$, which are fully consistent with our result. 

For the Bullet cluster, our mass estimate is rather poor due to a fairly low $n_\textrm{gal}$, but we can recover the result of \citet{Clowe04.1} within errors.
This comparison is, however, not as straightforward as it seems. The original ground-based \emph{VLT} data in \citet{Clowe04.1} had a field of view of only 7 arcmin, hence the radial range probed there is almost entirely excluded in our fit that starts at 3 arcmin. We therefore acknowledge the similarity of our mass estimates with the literature value, but do not consider this a powerful result. 

The situation is different for \sptw{}, where the shear profile is more regular and our mass estimate is better constrained. Our estimate is in excellent agreement with the weak-lensing analysis from \citet{Gruen13.2}. Our central value is about twice as high as the estimate from \citet{High10.1} based on optical richness. Another recent mass estimate from SZ and X-ray scaling relations by \citet{Reichardt13.1} of $M_{500\mathrm{c}} = 6.50\pm0.79\ h_{70}^{-1}\ 10^{14} \mathrm{M}_\odot$ is again fully consistent with our lensing estimate, which we derive as $M_{500\mathrm{c}} = 6.4^{+2.6}_{-2.2}\cdot 10^{14} \mathrm{M}_\odot$ by assuming an NFW profile with $c_{200\mathrm{c}}=3.4$ as measured from our lensing data.

We conclude this section with a test on the robustness of the mass estimate against uncertainties in the numerous calibrations we have employed. To assess the impact of the calibrations, we repeated the NFW-profile fitting \emph{without} the calibrations. The cluster-member contamination correction from \autoref{sec:cl_contamination} alone increases the mass estimates by less than 5\% as it only affects the galaxies within $\approx$5 arcmin, and our fits start at 3 arcmin. The \photoz{} recalibration from \autoref{sec:photoz_calib} yields a global boost of the lensing signal by 5-10\%. The biggest impact stems from the noise-bias correction (\autoref{sec:noise-bias}), which globally increases the inferred shear by $\approx$20\%.\footnote{This amount of noise-bias correction seems high by cosmic shear standards, but we also include much fainter galaxies with our adopted cuts. Similar levels of noise bias have been reported for various methods, e.g. by \citet[their section 5.5]{Massey07.1}.} The sum of all these calibrations amounts to a considerable 35\%, so that uncertainties in the calibrations actually become important. As we have laid out in the relevant sections, these calibrations are determined quite well with dedicated measurements, but we will conservatively allow for a 20\% systematic error budget. Compared to the statistical uncertainties that are of order 50\% (with \rxj{} being the only cluster with a 25\% statistical error), we conclude that the overall error is dominated by shape noise from the dispersion of galaxy ellipticities.
\medskip

\subsection{Richness-mass relation}
\label{sec:mass-richness}

An obvious additional cross-check for the data in this work is to compare it with the mass-richness relation for low-redshift clusters. \citet{rykoff+12} constrained it with maxBCG \citep{Koester07.1} clusters in the range $0.1\leq z \leq 0.3$ with a very similar richness estimator $\lambda$ to the one we employ here.
Although their redshift range only covers two of our clusters (the other two are at slightly higher redshift), we expect that deviations would more likely stem from our large measurement errors on the weak-lensing mass than from any possible redshift evolution of that relation.
 
We list the \redmapper-estimated richness and redshift estimates in \autoref{tab:results} and note that for the three clusters, for which we have spectroscopic redshifts, \redmapper{} provides excellent redshift estimates, with deviations within $2\Delta z_\lambda$ in all cases. We take this as an indication that our overall photometric calibration (\autoref{sec:photo_calib}) and the red-sequence (\autoref{sec:cl_sel}) calibration were successful. In \autoref{fig:mass_richness} we compare the richnesses with the weak-lensing masses from \autoref{fig:shear_profiles} and the best-fit solution from \citet[their Eq. B5]{rykoff+12}
\begin{equation}
\label{eq:mass-richness}
\ln \left(\frac{M_{200\mathrm{c}}}{10^{14} h^{-1} \mathrm{M}_\odot}\right) = 1.48 (1 \pm 0.33) + 1.06 \ln \left(\frac{\lambda}{60}\right)
\end{equation}
and find that our measurements indeed agree with the expectations,\footnote{We note that \citet{rykoff+12} made simplifying assumptions that entail e.g. the absence of an error on the slope in \autoref{eq:mass-richness}. We therefore refer to their Appendix B for a discussion of the limitation of the inferred mass-richness relation.} within the considerable scatter both measurements exhibit.

\begin{figure}
\includegraphics[width=\linewidth]{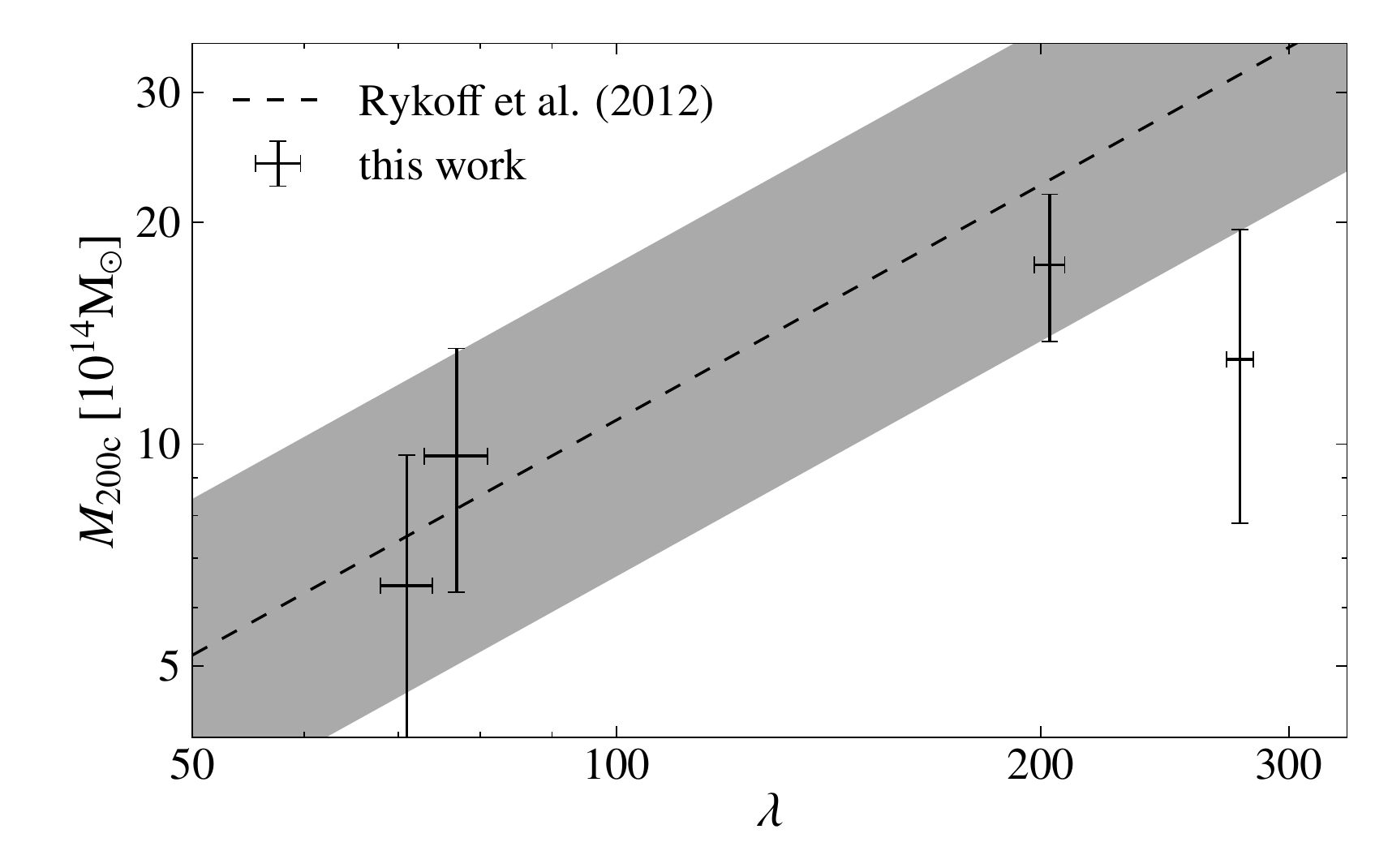}
\caption{Lensing mass $M_{200\mathrm{c}}$ as a function of \redmapper's richness $\lambda$ for the four investigated clusters. The \emph{dashed} line shows the expected scaling relation from \citet[see \autoref{eq:mass-richness}]{rykoff+12} with their proposed relative scatter of 33\% at fixed richness (\emph{shaded region}).}
\label{fig:mass_richness}
\end{figure}

\section{Mass  and cluster galaxy distributions}
\label{sec:maps}

We now move from spherically averaged masses to two-dimensional maps of the weak-lensing mass and the cluster galaxies. We have seen in \autoref{sec:masses} that the NFW profile is an acceptable fit to the measurements. In detail, that is not even expected as the NFW profile only describes the \emph{average} radial profile of dark matter halos in simulations, incapable of reproducing the complex structures massive clusters often exhibit \citep[e.g.][]{Merten11.1,Medezinski13.1}. We are particularly interested in the environment of these clusters, using DECam to follow the filamentary structures from which the clusters accrete out to distances normally not accessible to dedicated cluster-lensing studies on imagers with smaller fields of view.

We start the inspection of the cluster fields visually at the central $5\times5$ arcmin$^2$ of each cluster in the left column of \autoref{fig:maps}, where we can see the BCG and other bright cluster members. In all four clusters, we can see that several of the obvious cluster member galaxies tend to line up along one axis that coincides with the orientation of the BCG. This long-known tendency \citep{Sastry68.1, Carter80.1} is still not entirely understood, but a plausible scenario entails that accretion of satellite halos along filaments determines the cluster major axis, and the BCG orients itself accordingly \citep[e.g.][and references therein]{Hao11.1}.

\begin{figure*}
\includegraphics[width=\linewidth, trim=0 20 0 21]{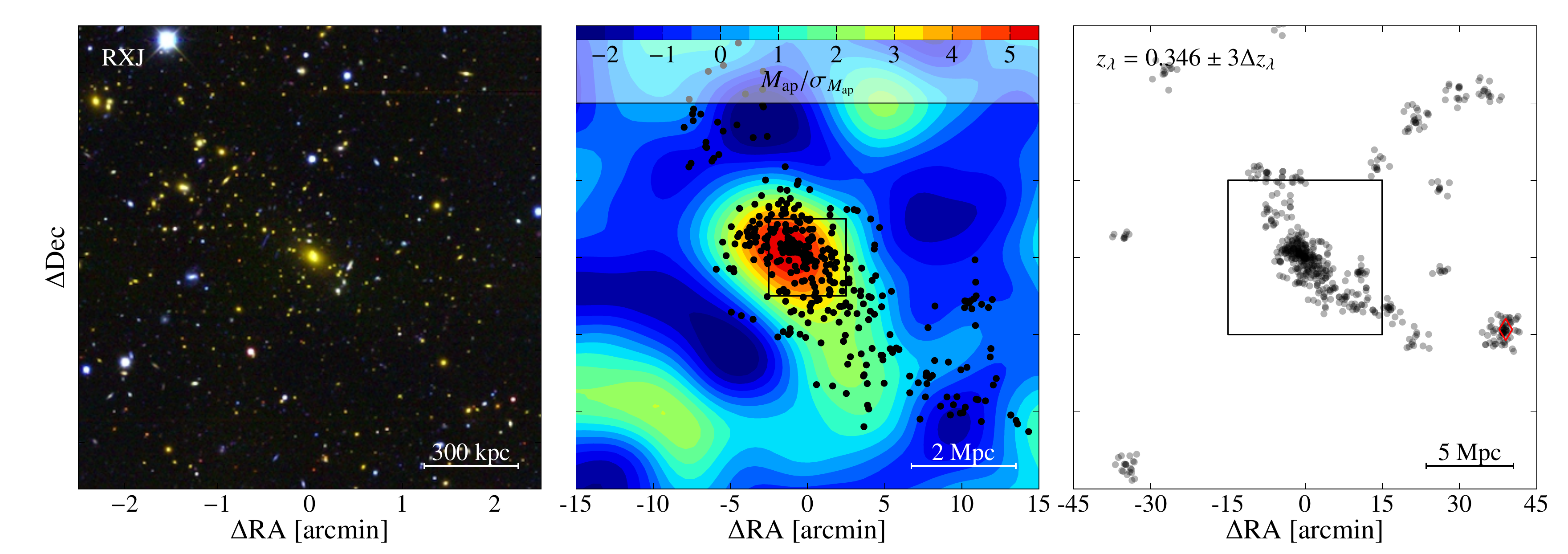}
\includegraphics[width=\linewidth, trim=0 20 0 21]{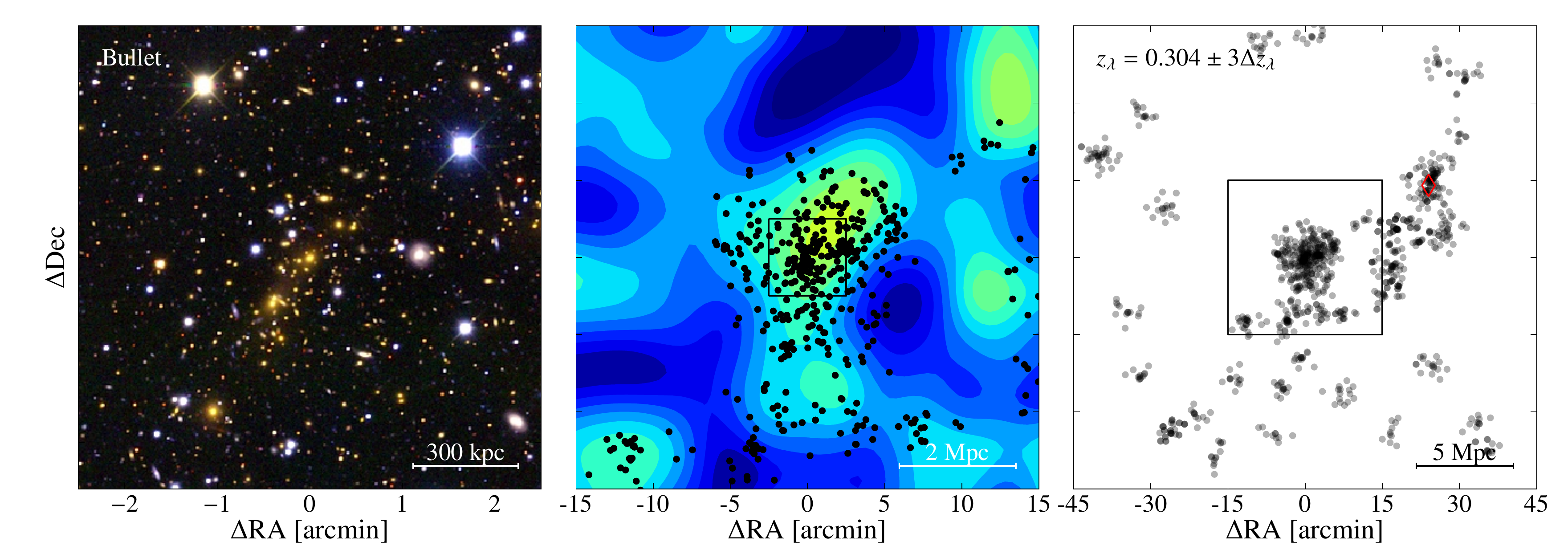}
\includegraphics[width=\linewidth, trim=0 20 0 21]{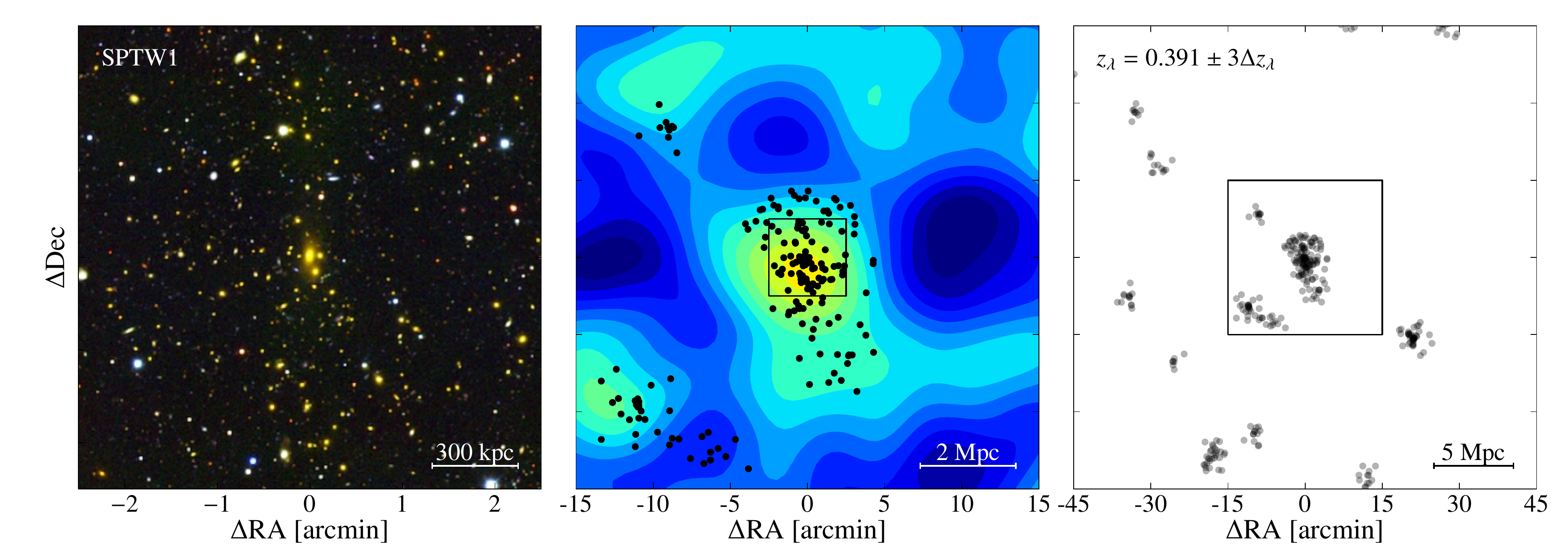}
\includegraphics[width=\linewidth, trim=0 0 0 21]{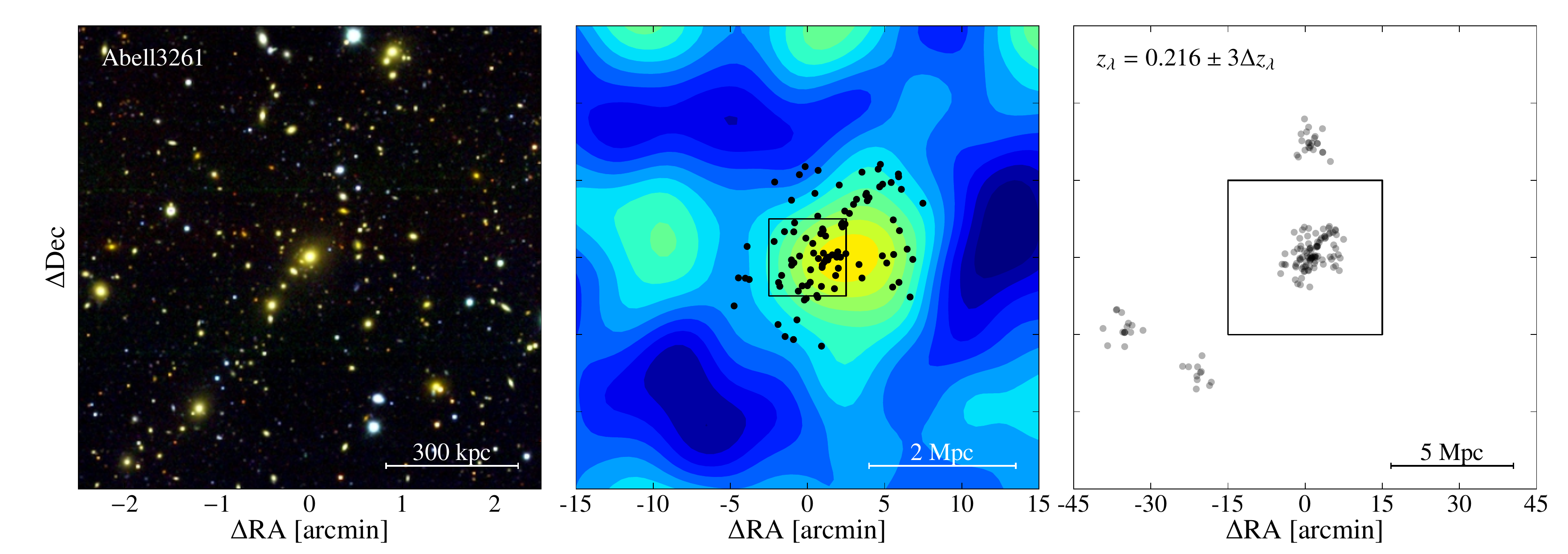}
\caption{\emph{1st column:} Multi-color image of the inner 5 arcmin.  \emph{2nd column:} Weak-lensing aperture mass significance map of the inner 30 arcmin (\emph{contours}, cf. \autoref{eq:M_ap}), overlaid with galaxies (\emph{black dots}) in \redmapper-detected groups with $\lambda\geq5$ and redshifts of $z_\lambda=z_\lambda^\textrm{c}\pm3\Delta z_\lambda$. \emph{3rd column:} The same \redmapper{} galaxies as in the 2nd column, but for the entire useable field of view of 90 arcmin. All panels are centered on the BCGs, the size of the previous (smaller) panel is indicated by black boxes in columns 2 and 3. From top to bottom: \rxj, \bulletcl, \sptw, and \abell.}
\label{fig:maps}
\end{figure*}

\subsection{Mass maps}
To perform the mass reconstruction of the galaxy clusters, we move further out to cover 30 arcmin, a scale typical of weak-lensing studies of individual galaxy clusters, and employ the aperture-mass technique from \cite{Schneider96.1}. It exploits that a local estimate of the convergence $\kappa(\theta)=\Sigma(\theta)/\Sigma_\textrm{crit}$ can be obtained by summing up all ellipticity measurements $\epsilon_\mathrm{t}(\theta_j)$ inside of a circular aperture,
\begin{equation}
\label{eq:M_ap}
M_\textrm{ap}(\theta) = \sum_j Q(|\theta - \theta_j|)\ \epsilon_\textrm{t}(\theta_j).
\end{equation}
Here the tangential component of the ellipticity $\epsilon_\mathrm{t}$ is calculated with respect to the center $\theta$ of the aperture, not the center of the cluster as in \autoref{eqn:likelihood}. Under the assumption of uncorrelated Gaussian noise in the ellipticities with variance $\sigma_\epsilon^2$, the variance of $M_\textrm{ap}$ is given by
\begin{equation}
\sigma^2_{M_\textrm{ap}} = \frac{1}{2}\sum_j Q^2(|\theta - \theta_j|)\ |\epsilon|^2(\theta_j).
\end{equation} 
So far we have not specified the weight function $Q$, and in fact we have considerable freedom in doing so, which allows us to demand additional desirable properties of the reconstructed mass maps. Since the noise contribution stemming from $\sigma_\epsilon$ is scale-free, the maximum $M_\textrm{ap}/\sigma_{M_\textrm{ap}}$ is achieved if $Q$ is identical to the signal we try to find, i.e. the tangential shear of the cluster \citep{Schneider96.1, Bartelmann01.1}. Thus, we could turn the measured shear profiles from \autoref{fig:shear_profiles} into templates for optimal individual shapes of $Q$, which would result in mass maps that are not easily comparable against each other. We therefore seek a common weight function shape $Q(r)$ with a single characteristic radius $R_\textrm{ap}$, knowing that we will sacrifice some statistical significance with this decision. 
We follow the design choices of \citet{Schneider96.1}, which we find particularly suitable for this work for three reasons. First, he approximated the shear profile by an isothermal $\gamma_\mathrm{t}\propto1/r$ relation, which should allow us to capture the extended environment of these massive, and in parts even merging, clusters better than the steeper NFW profile. Second, the weight function excises an inner circle at $r<\nu_1 R_\textrm{ap}$ to avoid regions where the relation between $\gamma$ and $\kappa$ becomes nonlinear and shape measurements are rendered difficult due to bright cluster members. The same concern has led us to exclude the inner regions when fitting the NFW profile in \autoref{sec:masses}. Third, \citet{Schneider96.1} also sets the outer edge of the weight function at $r>R_\textrm{ap}$ to allow the filter to operate on finite, and potentially masked, fields and to avoid the inclusion of truly uncorrelated structures, again corresponding to decisions we made earlier. To avoid a sharp cutoff at that outer edge, we let the filter roll off smoothly, starting at $\nu_2 R_\textrm{ap}$, where $\nu_2<1$. Considering the scale, over which we can find a noticeable shear signal in \autoref{fig:shear_profiles}, we choose $R_\textrm{ap}=10$ \mbox{arcmin}, with an inner exclusion region of $\nu_1 R_\textrm{ap}=1$ arcmin and the onset of the roll-off at $\nu_2 R_\textrm{ap} = 9$ arcmin. The exact form of the $Q(r)$ can be seen in \citet[eq. 34, Figs. 1 and 2]{Schneider96.1}.
It is worth pointing out that in adopting these choices we employ a filter that is substantially different from those that attempt to maximize purity of blind detections from wide-field weak-lensing data  \citep[e.g.][]{Maturi05.1} by suppressing the influence of large-scale structure fluctuations: here we know where the clusters are and we \emph{want} to probe the correlated material surrounding the clusters.

We present the resulting mass maps, i.e. maps of $M_\textrm{ap}/\sigma_{M_\textrm{ap}}$ centered on the location of the BCG, as contours in the middle panels of \autoref{fig:maps}. The mass maps of RXJ and SPTW1 show clearly significant peaks, exceeding 5.5 and 3.5$\sigma$ in their respective centers. For the Bullet cluster, the peak significance is not as prominent despite having an expected mass comparable to RXJ, but due to its highly non-spherical mass distribution, the spherical filter shape works against the signal, reducing its amplitude. Finally, even the least massive cluster, \abell, shows up at the level of $3\sigma$ in its mass map. The reduced significance of the latter two is caused also by a low $n_\textrm{gal}\approx6$ arcmin$^{-2}$ after background cuts. 

We overlay the mass maps with \redmapper-detected galaxies in groups with $\lambda \geq 5$, whose redshift estimates $z_\lambda$ are consistent within $\pm3\Delta z_\lambda \lesssim 0.03$ with the main cluster redshift $z_\lambda^\textrm{c}$ (see \autoref{sec:cl_sel} for details). The distribution reveals the structure of the red-sequence galaxies within and around the main cluster halo.

Several aspects of the mass maps are remarkable. First, the mass maps clearly follow the red-sequence cluster-galaxy distributions. This is additional confirmation that the shape measurements indeed perform well since we expect mass to trace light. Second, the peaks in the mass maps do not always coincide with the cluster BCGs. For the Bullet cluster, the peak is placed roughly between the main cluster and the subcluster. Given our large smoothing scale, it is not surprising that the two peaks are effectively merged. Furthermore, even for single-peaked mass distributions, such shifts between the BCG and the most prominent mass peak are a consequence of sampling the shear field with a finite number of sources \citep[e.g.][]{Dietrich12.1}. Third, we note that additional peaks exist in the mass maps, occasionally reaching $2\sigma$, that are not associated with the main cluster or other known clusters in the fields. To test the robustness of the mass maps we therefore run bootstrap resamples of the lensed galaxies, which reveal that these peaks are largely spurious and depend on particular configurations of a small number of neighboring galaxies. In contrast, the peaks associated with the main clusters shift location by up to 2 arcmin but are otherwise robust under resampling.

\subsection{Filamentary features}
\label{sec:filaments}

\begin{figure*}
\includegraphics[width=\linewidth]{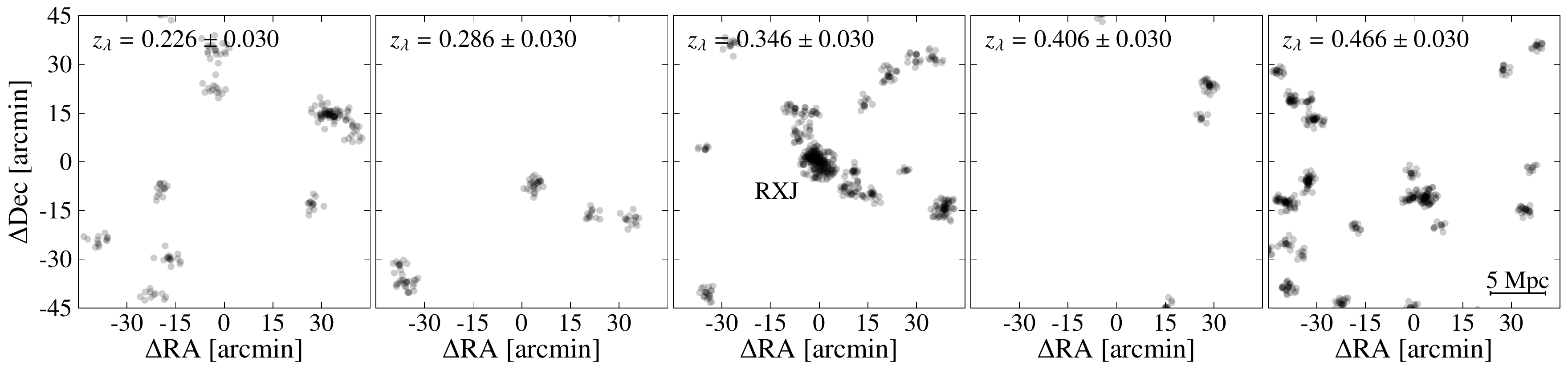}
\caption{Galaxies in \redmapper-detected groups in the field of \rxj{} with $\lambda\geq5$ and redshifts of $z_\lambda$ within the indicated non-overlapping redshift slices, centered on the \redmapper-assigned redshift $z_\lambda^\textrm{c}=0.344$ of the main cluster.}
\label{fig:redmapper_slices}
\end{figure*}

In the third column of \autoref{fig:maps}, we utilize the full DECam image and show the distribution of \redmapper-detected groups for the entire usable area. We can see that the most massive clusters in the sample show a rich environment that seems connected to the central region and that reaches out to other clusters in the vicinity. From a hierarchical CDM structure-formation scenario, we expect such structures, called filaments, to be attached to and to act as bridges between clusters, especially the very massive ones \citep[e.g.][]{Bond96.1}. Both cosmological simulations and spectroscopic surveys have revealed filaments with typical lengths of ten to dozens of Mpc \citep[e.g.][]{Alpaslan14.1, Tempel14.1}. Of particular relevance to this study is that in simulations more than 80\% of cluster pairs with distances of around 10 Mpc$/h$ are connected by filaments \citep{Colberg05.1}.
In the RXJ field, there is another cluster in the same redshift slice, detected by \redmapper{} with $\lambda=41\pm2$ (\emph{red diamond} in the top-right panel of \autoref{fig:maps}), at a distance of 40 arcmin to the south-west (12 Mpc in the plane of the sky), and a string of less massive groups that may constitute a mildly curved connecting filament. For the Bullet cluster, we can see another cluster at 28 arcmin to the north-west ($\lambda=78\pm3$) and other correlated structure over 20 Mpc, assuming the entire galaxy distribution is aligned in the plane of the sky -- a slight underestimation because it is known that at least the inner region is inclined by $10 - 15^\circ$ \citep{Barrena02.1, Markevitch02.1}. Structures that large have only been observed around a few other clusters, mostly at higher redshift than we probe here \citep[e.g.][]{Tanaka09.1, Verdugo12.1}. 

Without spectroscopic follow-up, we cannot prove that all shown \redmapper-detected galaxies are indeed at the redshift of the main cluster or gravitationally interacting with the main halo. There are, however, additional aspects that support the notion that the shown structures are indeed real and associated with the clusters. \redmapper{} determines the redshifts of the main cluster halos with high accuracy (compare \autoref{tab:clusters} with \autoref{tab:results}), which implies that the overall photometric calibration performs well and that the red-sequence colors are properly calibrated for the DES photometry. This is consistent with the results from Rykoff et~al. (in prep.) showing that groups with $\lambda\geq 5$ exhibit scatter of $\sigma_{z_\lambda} \leq 0.015$ and negligible bias when compared to existing spectroscopic redshifts of clusters in the DES SV footprint.
We can also split the light cone of the observed fields into thin redshift slices to test whether the structures are confined to the redshift of the main cluster halo. The result for the RXJ cluster is shown in \autoref{fig:redmapper_slices}. Note that this test is different from the right panels in \autoref{fig:maps} in that we do not ask whether the redshift of the group is consistent with the main cluster's at $3\Delta z_\lambda$ (of each group), which potentially allows for an arbitrarily wide redshift range if $\Delta z_\lambda \rightarrow\infty$. Instead, we only consider the central value $z_\lambda$ of each group and fix the width of the slice at $\pm0.03$, the typical value of $3\Delta z_\lambda$ for groups with $\lambda\geq 5$. This way the influence of chance projections of groups with poorly determined redshifts can be suppressed.  We can indeed see that the filamentary structure does not bleed into other slices. Also, apart from additional smaller clusters at different redshifts, the other slices are much less populated and do not show similarly prominent correlated structures. Tests of the other clusters yield similar results. 

\section{Summary}
\label{sec:discussion}

In the study presented here we observed four galaxy cluster fields with the newly installed imager DECam and tested all data processing stages necessary for weak-lensing applications within the Dark Energy Survey. Even with early data observed during the Science Verification phase, we find the instrument and these pipelines to perform according to anticipated specifications and yield astrometry, photometry, and shape measurements adequate for the pathfinder analysis presented here, with no show-stoppers that preclude forthcoming science analyses.
Most important in this work was to establish how to obtain reliable shape catalogs from DECam data, and we summarize our findings as follows:
\begin{itemize}[leftmargin=*]
\item By jointly fitting for the astrometry of all exposures, we find 20 mas scatter in the astrometric solution of \scamp{} across the entire focal plane.
\item The PSF patterns are spatially fairly smooth across the focal plane and can be well modeled with \psfex, provided that the brightest stars are discarded to limit the impact of the flux dependence of the PSF width. The majority of the coadd images have PSF model residuals in size and ellipticity that are subdominant compared to shape scatter up to separations of 1 degree.
\item With suitably chosen cuts, \shape{} yields shape measurements with a number density $n_\textrm{gal} \approx10$ arcmin$^{-2}$ as expected for full-depth data at nominal seeing. The results are consistent when varying source flux, size, \photoz{} or the filter of observation.
\end{itemize}
These technical prerequisites enable us to utilize the large field of view of DECam to estimate weak-lensing masses and to map out the galaxy and mass distribution of the targeted galaxy clusters. Our scientific results are:
\begin{itemize}[leftmargin=*]
\item We find weak-lensing masses for \rxj{}, the Bullet cluster \bulletcl{}, and \sptw{} that are in good agreement with previous studies. For clusters at higher redshift or dedicated high-precision lensing studies of individual systems, deeper imaging than the nominal DES depth of $10\times90$ seconds is advised.
\item For the cluster \abell, we provide the first redshift, richness, and weak-lensing mass estimates in the literature.
\item The mass maps of all four clusters show their most significant peak at or close to the cluster BCG. Clusters with a visibly noticeable alignment of cluster member galaxies exhibit the same orientation also in the mass maps.
\item Due to well-calibrated photometry, the red-sequence method \redmapper{} detects these four clusters reliably. The redshift estimates of even much smaller clusters with $\lambda > 5$ are precise enough (Rykoff et al., in prep.) to form thin slices at the cluster redshift and to map out the distribution of red-sequence galaxies in the entire cluster environment. The most massive systems, \bulletcl{} and \rxj, show filamentary structures over about 1 degree, equivalent to about 20 Mpc at the cluster redshifts. If the presence of these structures can be confirmed, this technique can be employed in the DES main survey and enable efficient searches for large-scale filaments in the entire DES footprint without the need for full spectroscopic coverage. 
\end{itemize}
The work presented here will form the basis of forthcoming analyses within DES, concerning e.g. the cluster-mass function, the calibration of mass-observable relations from optical richness, X-ray and SZE, and other more demanding lensing applications.

{\small
\section*{Acknowledgements}

We are grateful for the extraordinary contributions of our CTIO colleagues and the DES Camera, Commissioning and Science Verification teams in achieving the excellent instrument and telescope conditions that have made this work possible. The success of this project also relies critically on the expertise and dedication of the DES Data Management organization. 

Funding for the DES Projects has been provided by the U.S. Department of Energy, the U.S. National Science Foundation, the Ministry of Science and Education of Spain, the Science and Technology Facilities Council of the United Kingdom, the Higher Education Funding Council for England, the National Center for Supercomputing Applications at the University of Illinois at Urbana-Champaign, the Kavli Institute of Cosmological Physics at the University of Chicago, Financiadora de Estudos e Projetos, Funda{\c c}{\~a}o Carlos Chagas Filho de Amparo {\`a} Pesquisa do Estado do Rio de Janeiro, Conselho Nacional de Desenvolvimento Cient{\'i}fico e Tecnol{\'o}gico and the Minist{\'e}rio da Ci{\^e}ncia e Tecnologia, the Deutsche Forschungsgemeinschaft and the Collaborating Institutions in the Dark Energy Survey.

The Collaborating Institutions are Argonne National Laboratories, the University of California at Santa Cruz, the University of Cambridge, Centro de Investigaciones Energeticas, Medioambientales y Tecnologicas-Madrid, the University of Chicago, University College London, the DES-Brazil Consortium, the Eidgen{\"o}ssische Technische Hochschule (ETH) Z{\"u}rich, Fermi National Accelerator Laboratory, the University of Edinburgh, the University of Illinois at Urbana-Champaign, the Institut de Ciencies de l'Espai (IEEC/CSIC), the Institut de Fisica d'Altes Energies, the Lawrence Berkeley National Laboratory, the Ludwig-Maximilians Universit{\"a}t and the associated Excellence Cluster Universe, the University of Michigan, the National Optical Astronomy Observatory, the University of Nottingham, the Ohio State University, the University of Pennsylvania, the University of Portsmouth, SLAC National Accelerator Laboratory, Stanford University, the University of Sussex, and Texas A\&M University.

PM, ES, EH, KP, KH are supported by the U.S. Department of Energy under Contract No. DE- FG02-91ER40690. SB, TK, MH, JZ, BR acknowledge support from European Research Council in the form of a Starting Grant with number 240672. DG was supported by SFB-Transregio 33 ``The Dark Universe" by the Deutsche Forschungsgemeinschaft (DFG) and the DFG cluster of excellence ``Origin and Structure of the Universe". The DES participants from Spanish institutions are partially supported by MINECO under grants AYA2009-13936, AYA2012-39559, AYA2012-39620, and FPA2012-39684, which include FEDER funds from the European Union.

This paper has gone through internal review by the DES collaboration.

\bibliography{references}
}
\appendix
\section{PSF flux dependence}
\label{sec:brighter-wider}

The most important test for the PSF model is whether it can reproduce the sizes and ellipticities of observed stars in the field. When using the full range of stellar fluxes to inform the PSF model, we unfortunately register that the PSF width $s$ is overestimated for the majority of all stars (see left panel of \autoref{fig:brighter-wider}). This is a direct consequence of the flux-dependent charge registration in the DECam CCDs, for which we currently do not yet have a chip-level correction. There is furthermore a broadening of the stellar ellipticity residuals, foremost in the $\epsilon_1$ direction, the cause of which is not fully understood.

Irrespective of the actual mechanism at work, we can effectively reduce the impact of the flux dependency by excluding the brightest stars when computing the PSF model. We found that rejecting the brightest 3 magnitudes below saturation level, corresponding roughly to \verb|MAG_AUTO| $\in[15,18]$, allows for very accurate PSF models as determined by the diagnostics in \autoref{eq:tolerances} and exemplified in \autoref{fig:psfmodel}, where we used the fainter star selection for the same field as in \autoref{fig:brighter-wider}. While there will be a slight misestimate of the effective PSF a galaxy at our faintest magnitudes of $i\simeq24.5$ would encounter, the change in flux and therefore in size compared to our fainter star selection is too small to be detectable by the diagnostics employed in this paper.
As a practical consequence, we need to work with coadded images where the fainter stars can reliably be discriminated from galaxies.

\begin{figure}
\includegraphics[width=\linewidth]{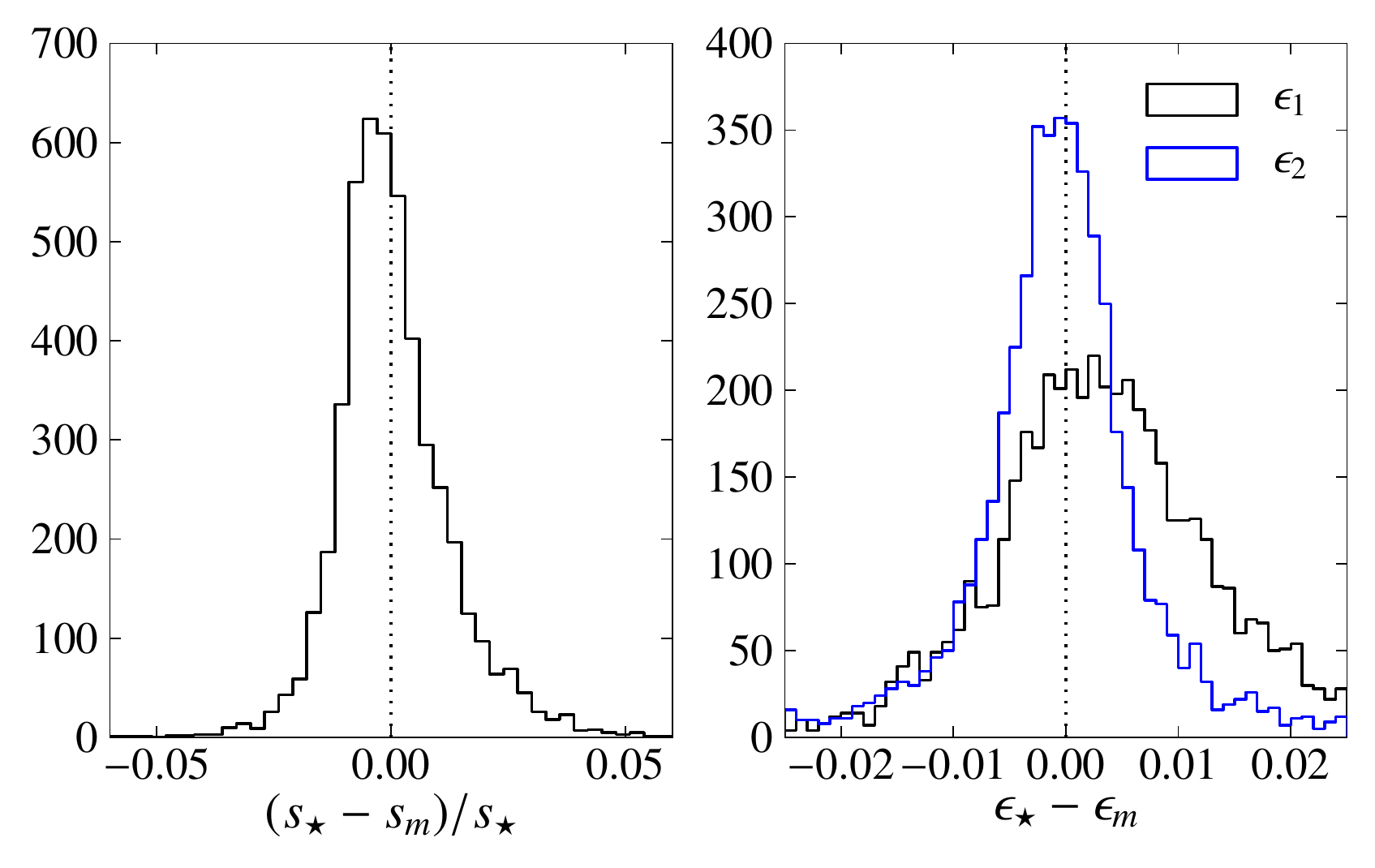}
\caption{Fractional size and ellipticity residuals of the \psfex{} model shown in \autoref{fig:psfmodel} but for a set of stars with \texttt{MAG\_AUTO} $\in [15, 21.5]$. The brighter stars in this selection lead the PSF model to adopt larger sizes and a preferred $\epsilon_1$-direction, which is not shared by the bulk of the (fainter) stars.}
\label{fig:brighter-wider}
\end{figure}

\section{Tolerances for PSF model diagnostics}
\label{sec:diagnostics}

In this section we seek to propagate failures of the PSF modeling approach in capturing both sizes and ellipticities of the actual PSF (and its spatial variation) into the the shear catalogs, thereby establishing limits on the required accuracy of the PSF models. 
We start with Eq. 13 from \citet{Paulin-Henriksson08.1},
\begin{equation}
\Delta\epsilon_{\rm sys} \simeq (\epsilon_{\rm gal} - \epsilon_\star) \frac{\Delta(s^2)}{s_{\rm gal}^2} - \Bigl(\frac{s_\star}{s_{\rm gal}}\Bigr)^2\Delta\epsilon,
\end{equation}
which estimates the systematic error in the shape of a deconvolved galaxy from the uncertainties in size and ellipticity of the PSF model, $\Delta(s^2)$ and $\Delta\epsilon$.\footnote{While the derivation strictly applies to moment-based measures only, and we seek to use it in a model-fitting context, the results depend weakly on the method used for shear estimation, provided the method yields an shape estimate that transforms as an ellipticity under an applied shear.}
Forming the correlation function yields
\begin{equation}
\label{eq:PSFerror}
\begin{split}
\langle \conj{\Delta\epsilon_{\rm sys}} \Delta\epsilon_{\rm sys}\rangle = &\big[\sigma_\epsilon^2 + \langle \conj{\epsilon_\star} \epsilon_\star\rangle\big] \Bigl\langle\Bigl(\frac{\Delta s^2}{s_{\rm gal}^2}\Bigr)^2\Bigr\rangle + \Bigl\langle\Bigl(\frac{s_\star}{s_{\rm gal}}\Bigr)^4\Bigr\rangle \langle \conj{\Delta\epsilon} \Delta\epsilon\rangle\\
 &+ \Bigl\langle\frac{\Delta s^2 s_\star^2}{s_{\rm gal}^4}\bigl(\conj{\epsilon_\star} \Delta\epsilon + \conj{\Delta\epsilon} \epsilon_\star\bigr)\Bigr\rangle.
 \end{split}
\end{equation}
Now we will have to make several assumptions to relate the terms arising here to the PSF model diagnostics defined in \autoref{eq:D12} and \autoref{eq:H1}. First, to pull out a common size-ratio for the first two terms, we need to assume that the size residuals do not correlate with stellar size:
\begin{equation}
\begin{split}
\langle \conj{\Delta\epsilon_{\rm sys}} \Delta\epsilon_{\rm sys}\rangle = &\Bigl\langle\Bigl(\frac{s_\star}{s_{\rm gal}}\Bigr)^4\Bigr\rangle\ \Bigl[\big[\sigma_\epsilon^2 + \langle \conj{\epsilon_\star} \epsilon_\star\rangle\big]\ \rho_3 + \rho_1\Bigr]\\
 &+ \Bigl\langle\frac{\Delta s^2 s_\star^2}{s_{\rm gal}^4}\bigl(\conj{\epsilon_\star} \Delta\epsilon + \conj{\Delta\epsilon}\epsilon_\star\bigr)\Bigr\rangle.
 \end{split}
\end{equation}
If we furthermote assume that $\Delta\epsilon$ and $\epsilon_\star$ as well as $\Delta s^2$ and $\Delta \epsilon$ are uncorrelated, we can simplify the last term to
\begin{equation}
\Bigl\langle\frac{\Delta s^2 s_\star^2}{s_{\rm gal}^4}\Bigr\rangle \langle\conj{\epsilon_\star} \Delta\epsilon + \conj{\Delta\epsilon}\epsilon_\star\rangle,
\end{equation} 
which allows us to identify it with $\rho_2$:
\begin{equation}
\label{eq:PSFDiagnosticLimits}
\langle\conj{\Delta\epsilon_{\rm sys}} \Delta\epsilon_{\rm sys}\rangle = \Bigl\langle\Bigl(\frac{s_\star}{s_{\rm gal}}\Bigr)^4\Bigr\rangle\ \Bigl[\big[\sigma_\epsilon^2 + \langle\conj{\epsilon_\star} \epsilon_\star\rangle\big]\ \rho_3 + \rho_1 + \Bigl\langle\frac{\Delta s^2}{s_\star^2}\Bigr\rangle\ \rho_2\Bigr].
\end{equation}
We note that these additional assumptions are clearly problematic as one could easily imagine  that residuals increase when the quantity that is modeled increases. For instance, both sizes and ellipticities tend to rapidly rise towards the field edges, where only a small number of stars can constrain the PSF model, a situation that should result in larger and correlated residuals for size and ellipticity.

Ideally, one would assess PSF model errors directly from \autoref{eq:PSFerror}, which considers all possible correlation between sizes and ellipticities (and their errors). We leave this to a forthcoming investigation and want to highlight another immediate consequence of our derivation. If we accept the limitations laid out above resulting in \autoref{eq:PSFDiagnosticLimits}, we see that for $\rho_3$ a prefactor of order $10^{-2}$ and for $\rho_2$ of order $10^{-3}$ (for a reasonably well-fit PSF model) reduces their relative impact on the total shape error. In other words, if all diagnostic correlation functions were equal, $\rho_1$ is most demanding, followed by $\rho_3$ and then $\rho_2$. In practice, we find relatively larger size than ellipticity residuals, rendering $\rho_3$ a useful and, in the case of the flux-dependent PSF, even decisive diagnostic. On the other hand, due to its very small pre-factor, $\rho_2$ appears not to carry substantial information to assess the PSF model quality. We will therefore refrain from enforcing limits on $\rho_2$ and will assess the PSF models with $\rho_1$ and $\rho_3$ only.

The LHS of \autoref{eq:PSFDiagnosticLimits} differs from the shear estimate only by the shear responsivity $P_\gamma$, which allows us to compare the total systematic budget with the statistical limit from the intrinsic shape scatter of the galaxies. For the two-point function, the number of galaxy pairs in a distance bin around $r$ is given by $n_{\rm gal} \pi (R_{\rm max}^2 - R_{\rm min}^2)$, with $R_{\rm min/max}$ denoting the minimum and maximum radius of that bin. Assuming a Gaussian form of the intrinsic shape scatter with variance $\sigma_\epsilon^2$, we get
 \begin{equation}
\label{eq:sys_limit}
P_\gamma^{-2}\langle\conj{\Delta\epsilon_{\rm sys}} \Delta\epsilon_{\rm sys}\rangle < \frac{\sigma_\epsilon^2}{n_{\rm gal} \pi (R_{\rm max}^2 - R_{\rm min}^2)}.
\end{equation}
We still need an estimate for the pre-seeing size of galaxies $s_{\rm gal}$ in our shape catalogs. For Gaussian-shaped galaxies and stars, one can directly relate the measurement of \verb|FWHM_RATIO| from \shape{} to the ratio of the moment-based size definition $s$ we have adopted in this paper:
\begin{equation}
\frac{s_{\rm gal}^2}{s_\star^2} = \verb|FWHM_RATIO|^2 - 1.
\end{equation}
Together with the shear responsivity yields
\begin{equation}
\label{eq:T}
T = P_\gamma \frac{s_{\rm gal}^2}{s_\star^2} = P_\gamma (\verb|FWHM_RATIO|^2 - 1).
\end{equation}
Finally, requiring that no diagnostic function alone crosses the limit set by \autoref{eq:sys_limit}, we get the set of tolerances in \autoref{eq:tolerances}.

\section*{Affiliations}
{\small
\begin{enumerate}[label=$^{\arabic*}\,$, leftmargin=*, align=left]
\item Center for Cosmology and Astro-Particle Physics, The Ohio State University, Columbus, OH 43210, USA
\item Department of Physics, The Ohio State University, Columbus, OH 43210, USA
\item Department of Physics \& Astronomy, University College London, Gower Street, London, WC1E 6BT, UK
\item Jodrell Bank Center for Astrophysics, School of Physics and Astronomy, University of Manchester, Oxford Road, Manchester, M13 9PL, UK
\item SLAC National Accelerator Laboratory, Menlo Park, CA 94025, USA
\item University Observatory Munich, Scheinerstrasse 1, 81679 Munich, Germany
\item Max Planck Institute for Extraterrestrial Physics, Giessenbachstrasse, 85748 Garching, Germany
\item Department of Physics and Astronomy, University of Pennsylvania, Philadelphia, PA 19104, USA
\item Institute of Cosmology \& Gravitation, University of Portsmouth, Portsmouth, PO1 3FX, UK
\item Kavli Institute for Cosmological Physics, University of Chicago, Chicago, IL 60637, USA
\item Fermi National Accelerator Laboratory, P. O. Box 500, Batavia, IL 60510, USA
\item Brookhaven National Laboratory, Bldg 510, Upton, NY 11973, USA
\item Cerro Tololo Inter-American Observatory, National Optical Astronomy Observatory, Casilla 603, La Serena, Chile
\item Space Telescope Science Institute, 3700 San Martin Drive, Baltimore, MD  21218, USA
\item Argonne National Laboratory, 9700 South Cass Avenue, Lemont, IL 60439, USA
\item Carnegie Observatories, 813 Santa Barbara St., Pasadena, CA 91101, USA
\item Institut d'Astrophysique de Paris, Univ. Pierre et Marie Curie \& CNRS UMR7095, F-75014 Paris, France
\item Kavli Institute for Particle Astrophysics \& Cosmology, 452 Lomita Mall, Stanford University, Stanford, CA 94305, USA
\item Institut de Ci\`encies de l'Espai, IEEC-CSIC, Campus UAB, Facultat de Ci\`encies, Torre C5 par-2, 08193 Bellaterra, Barcelona, Spain
\item Observat\'orio Nacional, Rua Gal. Jos\'e Cristino 77, Rio de Janeiro, RJ - 20921-400, Brazil
\item Laborat\'orio Interinstitucional de e-Astronomia - LIneA, Rua Gal. Jos\'e Cristino 77, Rio de Janeiro, RJ - 20921-400, Brazil
\item George P. and Cynthia Woods Mitchell Institute for Fundamental Physics and Astronomy, and Department of Physics and Astronomy, Texas A\&M University, College Station, TX 77843,  USA
\item Department of Physics, Ludwig-Maximilians-Universit\"{a}t, Scheinerstr.\ 1, 81679 M\"{u}nchen, Germany
\item Excellence Cluster Universe, Boltzmannstr.\ 2, 85748 Garching, Germany
\item Department of Physics, University of Michigan, Ann Arbor, MI 48109, USA
\item Department of Astronomy, University of Michigan, Ann Arbor, MI 48109, USA
\item Institut de F\'{\i}sica d'Altes Energies, Universitat Aut\`onoma de Barcelona, E-08193 Bellaterra, Barcelona, Spain
\item Department of Astronomy, University of Illinois,1002 W. Green Street, Urbana, IL 61801, USA
\item National Center for Supercomputing Applications, 1205 West Clark St., Urbana, IL 61801, USA
\item Department of Physics, University of Illinois, 1110 W. Green St., Urbana, IL 61801, USA
\item Australian Astronomical Observatory, North Ryde, NSW 2113, Australia
\item ICRA, Centro Brasileiro de Pesquisas F\'isicas, Rua Dr. Xavier Sigaud 150, CEP 22290-180, Rio de Janeiro, RJ, Brazil
\item Instituci\'o Catalana de Recerca i Estudis Avan\c{c}ats, E-08010 Barcelona, Spain
\item Lawrence Berkeley National Laboratory, 1 Cyclotron Road, Berkeley, CA 94720, USA
\item Centro de Investigaciones Energ\'eticas, Medioambientales y Tecnol\'ogicas (CIEMAT), Madrid, Spain
\item Instituto de F\'\i sica, UFRGS, Caixa Postal 15051, Porto Alegre, RS - 91501-970, Brazil
\item SEPnet, South East Physics Network, UK (www.sepnet.ac.uk)
\end{enumerate}
}
\label{lastpage}
\end{document}